\documentclass[jcp,twocolumn,unsortedaddress]{revtex4-1}
\usepackage{graphicx}
\usepackage{amsmath}
\usepackage{amssymb} 
\usepackage[usenames, dvipsnames]{color}

\usepackage[utf8]{inputenc}
\usepackage[english]{babel}
\usepackage{epstopdf}
\usepackage{hyperref}
\usepackage{braket}
\usepackage{float}
\usepackage{placeins}
\usepackage[overload]{empheq}
\usepackage{cases}
\usepackage[version=4]{mhchem}
\usepackage{mathrsfs}
\usepackage{wrapfig}

\newcommand{\pardd}[2]{\frac{\partial ^2 {#1}}{\partial {#2} ^2}}

\newcommand{\lB}{l_\mathrm{B}}

\newcommand{\Epar}{E_\parallel}

\newcommand{\e}{\mathrm{e}}
\newcommand{\dtilde}{\widetilde{d}}
\newcommand{\permr}{\varepsilon_{r}}

\newcommand{\kB}{k_\mathrm{B}}
\newcommand{\Fpar}{F_\parallel}
\newcommand{\utilde}{\widetilde{u}}
\newcommand{\umax}{\utilde_\mathrm{m}}
\newcommand{\Qtilde}{\widetilde{Q}}

\graphicspath{{./images/}}

\begin{document}

\title{Electroosmosis as a probe for electrostatic correlations}

\author{Ivan Palaia}
\affiliation{Department of Physics and Astronomy, Institute for the Physics of Living Systems
University College London, London WC1E 6BT, UK}
\affiliation{MRC Laboratory for Molecular Cell Biology, University College London, London WC1E 6BT, UK}

\author{Igor M. Telles}
\affiliation{Instituto de F\'isica, Universidade Federal do Rio Grande do Sul, Caixa Postal 15051, CEP 91501-970, Porto Alegre, RS, Brazil.}

\author{Alexandre P. dos Santos}
\affiliation{Instituto de F\'isica, Universidade Federal do Rio Grande do Sul, Caixa Postal 15051, CEP 91501-970, Porto Alegre, RS, Brazil.}

\author{Emmanuel Trizac}
\affiliation{Universit\'e Paris-Saclay, CNRS, LPTMS, 91405 Orsay, France.}

\begin{abstract}
We study the role of ionic correlations on the electroosmotic flow in planar double-slit channels, without salt. We propose an analytical theory, based on recent advances in the understanding of correlated systems. We compare the theory with mean-field results and validate it by means of dissipative particle dynamics simulations. Interestingly, for some surface separations, correlated systems exhibit a larger flow than predicted by mean-field. We conclude that the electroosmotic properties of a charged system can be used, in general, to infer and weight the importance of electrostatic correlations therein.
\end{abstract}
 
\maketitle


\section{Introduction}

The proof that natural colloids bear a surface charge in aqueous conditions dates back to 1809, with the electrokinetic experiments of Reus: he observed electrophoresis of clays, together with electroosmosis of water through sand~\cite{RusselSaville}. These two effects, driven by an electric field, can be viewed as describing the same phenomenon in different frames: while electrophoresis refers to the motion of charged macromolecules in a fluid at rest, electroosmosis is for the displacement of liquid when a solid interface (such as a capillary) is fixed~\cite{Hunter,Bruus}.
The external electric field applies a force on the ions of the electric double layer, that in turn transfer it to the surrounding solvent by means of viscous interactions. Applications abound in micro or nanofluidics, from pumping devices to blue energy and biological systems~\cite{BocquetRev2010,Kirby,Marbach2019}.

While the validity of hydrodynamic continuum approaches is challenged by downsizing~\cite{BocquetRev2010}, electrokinetic phenomena at small scale
are usually described at a continuum mean-field~(MF) level
when it comes to Coulombic effects, with neglect of ionic correlations~\cite{RusselSaville,Hunter,Marbach2019,Kirby,Anderson,BocquetRev2010,KiNe05,NiGr06,WaSh08,SaVa17,Bruus}. Due account of these effects has prompted only few theoretical attempts~\cite{Bazant2009,Storey2012}.
In the vast body of computational simulations as well, 
from lattice-Boltzmann approaches~\cite{SmSe09,WaKa10,RoPa13} to explicit water molecular dynamics simulations~\cite{KiDa06,ReAz18} or dissipative particle dynamics~(DPD)~\cite{SmSe09,JaZa18}, the question remains somewhat overlooked.

The study of ionic correlations, that eludes the Poisson-Boltzmann~(PB) framework, has focused on static systems~\cite{Le02}. 
Different types of approaches have been put forward, from field theoretic~\cite{Netz2001,Santangelo2006,Kanduc2008,Naji2013} to more heuristic works~\cite{RoBl96,Shklovskii1999PRL,Le02,Samaj2011,Samaj2018a}. We proceed along the latter angle of attack, to address the question of electroosmotic transport. We make use of the recent approach developed in~\cite{Samaj2018a}  to present a theory of electroosmosis in salt-free systems, that naturally includes Coulombic correlations.
Explicit dynamical quantities can be worked out, and we present as well DPD simulations that validate the theoretical analysis.

\begin{figure}
\begin{center}
\centering
\includegraphics[width=\columnwidth]{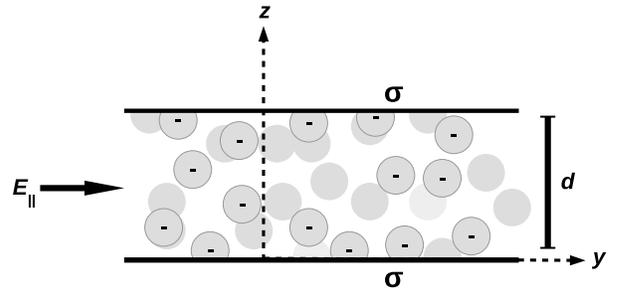}
\caption{Sketch of the system. Counterions are confined between two charged planar surfaces at distance $d$. An external electric field $\Epar $ is applied in the direction parallel to the surfaces.}
\label{fig1}
\end{center}
\end{figure}

We study in particular the double-slit geometry represented in Fig.~\ref{fig1}: two parallel planar surfaces of surface charge density $\sigma$ are placed at distance $d$ from each other and delimit a solution containing ions of opposite charge. In Sec.~\ref{sec:Theory} we solve the electroosmotic profile in the PB MF limit and in the correlated regime. In Sec.~\ref{sec:Sim} we compare theory and DPD simulations, showing ionic density profiles, velocity profiles and integrated flow. This comparison validates the theory in the strongly correlated regime and defines its limits of application for weekly correlated systems. Results are discussed in Sec.~\ref{sec:Conclusion}. We will show that there is a range of widths $d$, for which correlations enhance electroosmosis,
when compared to the MF expectation. This is an interesting phenomenon, that can play a role in situations of high confinement (e.g.~nano-/subnanofluidic devices, membrane porins), where solvent destructuring close to the charged surfaces can boost ionic correlations~\cite{CementInPrep, Schlaich2019},
with yet unexplored effects on the electroosmotic flow.

\section{Theoretical model}
\label{sec:Theory}

The solvent is assumed to be a structureless dielectric medium with relative permittivity $\permr$ and dynamic viscosity $\eta$. Under steady state conditions, the velocity profile of the fluid $u(z)$ obeys the following Stokes equation:
\begin{equation}
\label{eq:EOFss}
\pardd{u}{z}=- \frac{q e E_\parallel}{\eta} n(z)\,.
\end{equation} 
where $n(z)$ is the average density of counterions, $q$ their valence, and $e$ the elementary charge. The solvent obeys no-slip boundary conditions
\begin{equation}
u(0)=u(d)=0\,.
\label{eq:noslip}
\end{equation}
When substituting the velocity field $u$ for the electric potential $\phi$,
Eq.~\eqref{eq:EOFss} yields the Poisson equation;  $\phi$ and $u$ are thus linearly related~\cite{Hunter}. A second key observation which stems from the planar geometry, where the fluid flows parallel to the
plates and thus perpendicular to the charged interface, is that ion migration does not influence the equilibrium density profile\cite{Hunter}.
Characterising the flow thus simply requires 
the knowledge of the equilibrium behaviour of the ionic density: in this sense, the ionic flow is enslaved to the equilibrium density profile.
We solve below the Stokes equation using for the (equilibrium) ionic density profile $n(z)$ either the MF (uncorrelated) form, or the strong coupling~(SC) (correlated) regime. 

\subsection{Mean-field regime}
In MF, the density of counterions is~\cite{Andelman2006}
\begin{equation}
\label{eq:densityPB}
n(z)= \frac{\sigma K}{q \tan \frac{K d}{2}} \frac{1}{\cos^2\left(K\left(z-\frac{d}{2}\right)\right)}\,.
\end{equation}
$K$ is given by the solution of the following equation
\begin{equation}
\label{eq:K}
K d \tan \frac{K d}{2} = \frac{d}{\mu}\,,
\end{equation}
where $\mu=(2\pi\lB q \sigma)^{-1}$ is the Gouy-Chapman length, $\lB=e^2/(4\pi\varepsilon_0\varepsilon_r \kB T)$ is the Bjerrum length and $T$ is temperature.

Plugging Eq.~\eqref{eq:densityPB} into Eq.~\eqref{eq:EOFss} and imposing conditions \eqref{eq:noslip}, one gets the steady-state electroosmotic profile in MF. Defining the convenient 
dimensionless velocity $\utilde$
\begin{equation}
    u(z)=\frac{e \Epar}{2\pi\lB \eta q} \utilde(z)\,,
    \label{eq:uunit}
\end{equation}
one gets
\begin{equation}
\label{eq:EOFPB}
\utilde(z)=\ln \frac{\cos\left(K\left(z-\frac{d}{2}\right)\right)}{\cos\frac{Kd}{2}}.
\end{equation} 
By symmetry, the flow is maximum at $z=\frac{d}{2}$. With $\umax=\utilde(\frac{d}{2})$, we have
\begin{equation}
\umax = -\ln \cos \frac{Kd}{2} \simeq \begin{dcases}
\frac{d}{4\mu}  & \text{if}\  d\ll\mu\\
\ln \frac{d}{\pi \mu} & \text{if}\  d\gg\mu
\end{dcases}\,.
\label{eq:EOFPBasymptotic}
\end{equation}
If $d\gg\mu$, $\umax$ diverges logarithmically with the distance, as does the electrostatic potential $\phi$. This is the fingerprint  of
rather inefficient screening by counterions only, as opposed to
systems where both types of microions are present, cations and anions
(added salt)~\cite{Andelman2006}. 
Indeed, the equilibrium density ionic profile decays like a power-law ($z^{-2}$) far from a single charged plate, as opposed to exponentially with salt. This corresponds to a slower convergence 
to neutrality, and there is thus a residual charge on which the applied electric field pulls. There is no similar mechanism at work under SC. Yet, it is possible to observe an enhanced flow at intermediate interplate distances, as we will discuss.

By integrating Eq.~\eqref{eq:EOFPB} for $z$ between $0$ and $d$, we get $Q$, the volume flow rate of fluid per unit width length (with dimensions of surface divided by time). In the following units
\begin{equation}
    Q=\frac{e\Epar \mu}{2\pi \lB \eta q} \Qtilde\,,
    \label{eq:Qunit}
\end{equation}
this is:
\begin{equation}
    \Qtilde= \frac{\mathrm{Cl}_2(\pi-Kd)-\mathrm{Cl}_2(\pi+Kd)}{2K\mu} -\frac{d}{\mu} \ln\left(2\cos\frac{Kd}{2}\right) \,,
    \label{eq:QPB}
\end{equation}
where $\mathrm{Cl}_2(x)=-\int_0^x \ln\left(2\sin\frac{t}{2}\right)\, \mathrm{d}t = \frac{i}{2}[\mathrm{Li}_2(e^{-ix})-\mathrm{Li}_2(e^{ix})]$ is the Clausen integral, or Clausen function of second order, or log-sine integral, and $\mathrm{Li}_2$ represents the dilogarithm function. $\Qtilde$ goes as $d^2$ for $d\ll\mu$, and as $d\ln(d)$ for $d\gg\mu$. This reflects the behaviour of the maximum velocity $\umax$, from Eq.~\eqref{eq:EOFPBasymptotic}.

\subsection{Strong coupling analysis}

We quantify the importance of ionic correlations by means of the so-called electrostatic coupling parameter~\cite{Moreira2000, Netz2000EPJE}, as routinely used in plasma physics:
\begin{equation}
\Xi=\frac{q^2\lB}{\mu}=2\pi q^3 \lB^2 \sigma\,.
\end{equation}
 The MF regime corresponds to $\Xi\to0$ (low charges, high permittivity), whereas higher $\Xi$ values correspond to more energy-dominated systems, with $\Xi\to\infty$ representing the ground state (formally, the behaviour at $T=0$, assuming $\varepsilon_r$ does not depend on $T$). In practice, deviations from PB become appreciable when $\Xi$ exceeds a couple of units. At room temperature in water ($\varepsilon_r=80$), this means $\sigma>1/q^3 \,$nm$^{-2}$. With monovalent ions $q=1$, this 
 condition is hardly met in practice, while it becomes more standard
when $q\geq 2$, with divalent or trivalent counterions.

As proposed in~\cite{Samaj2018a}, the density of counterions can be approximated by introducing an effective field $\kappa$. In the ground state, this is the total electric field felt by an ion leaned to a charged surface: this includes the contributions of both charged walls and, most importantly, of all the other ions. In practice it is defined by the energetic cost $\kappa \Delta z$ of moving such ion away from the surface by a small distance $\Delta z$. In the ground state ($\Xi\to\infty$), $\kappa$ can be computed analytically and varies with $d$ on a scale $a=\sqrt{q/\sigma}\propto\mu\sqrt{\Xi}$, the typical ``in-plane'' distance between ions. At finite coupling $\Xi$, a situation where fluctuations are present, this picture can be retained with very satisfactory results, as long as $\Xi$ stays $\gg1$~\cite{Samaj2018a}; the effective field, that from now on we express in units of $\kB T/\mu$, can be approximated by the following expression:
\begin{equation}
\label{eq:kappach1}
\kappa\left(\frac{d}{a}\right)= \frac{\frac{d}{a}}{\sqrt{\left(\frac{d}{a}\right)^2 + \frac{1}{2\pi}}}\,.
\end{equation}
Note that for $d\to 0$, ions are delocalised along $z$ and $\kappa$ goes to 0; for $d\to\infty$, one of the walls together with half of the ions are too distant to matter and $\kappa$ goes to 1, the electric field of a bare wall in our units.

The effective field is a useful tool to determine the average ion density, that can be written~\cite{Samaj2018a} as
\begin{equation}
\label{eq:SCdensity}
n(z)= \frac{2\pi\lB\sigma^2\kappa}{1-\e^{-\kappa d/\mu}} \left(\e^{\kappa \frac{z-d}{\mu}}+ \e^{-\kappa\frac{z}{\mu}} \right)\,.
\end{equation}
The solution of Eqs.~\eqref{eq:EOFss} and \eqref{eq:noslip} for density \eqref{eq:SCdensity} is
\begin{equation}
\label{eq:EOFscz}
\utilde(z)= \frac{1+\e^{-\kappa\frac{d}{\mu}}-\e^{\kappa\frac{z-d}{\mu}}-\e^{-\kappa\frac{z}{\mu}}}{\kappa\left(1-\e^{-\kappa\frac{d}{\mu}}\right)}\,,
\end{equation}
where we used the same units as in Eq.~\eqref{eq:uunit}.

\begin{figure}
\centering
\includegraphics[width=\columnwidth]{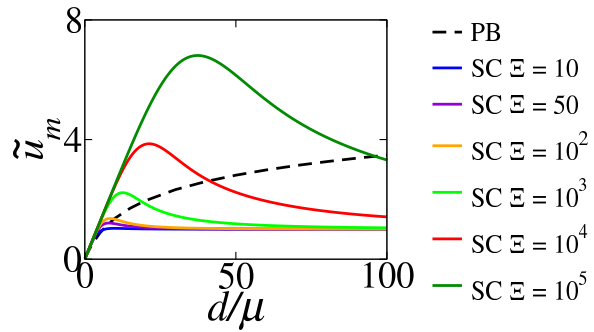}
\caption{
Maximum fluid speed $\umax$ as a function of dimensionless distance $d/\mu$. 
MF (PB) is represented by a black dashed line and SC by solid lines. To get the MF curve,
Eq.~\eqref{eq:K} is solved numerically and its result plugged into Eq.~\eqref{eq:EOFPB}. To get the SC curves, Eq.~\eqref{eq:EOFsc} is closed with formula \eqref{eq:kappach1}. 
}
\label{fig:plotbump}
\end{figure}

At the midpoint $z=d/2$, solvent flows with the maximum velocity
\begin{equation}
\label{eq:EOFsc}
\umax=\frac{\left(1-\e^{-\kappa\frac{d}{2\mu}} \right)^2}{\kappa\left(1-\e^{-\kappa\frac{d}{\mu}}\right)}\,.
\end{equation}
Asymptotically, this gives
\begin{equation}
\label{eq:EOFscAsymptotic}
\umax\simeq \begin{dcases}
\frac{d}{4\mu}  & \text{if}\  d \ll \mu\Xi^{1/4}\\
1 & \text{if}\  d \gtrsim \mu\Xi^{1/2}
\end{dcases}\,.
\end{equation}

In Fig.~\ref{fig:plotbump}, $\umax$ is plotted as a function of $d/\mu$, for different values of $\Xi$. Interestingly, $\umax$ is non-monotonic: it increases linearly (faster than the MF logarithmic solution) until a distance $\sim \mu\Xi^{1/4}$, then it peaks, decays (recrossing the MF curve), and eventually sets to a plateau for distances larger than $a\propto\mu\Xi^{1/2}$. The initial correlation-driven boost of the electroosmotic velocity, compared to MF, is due to the fact that correlations are more efficient than MF at delocalising ions at small distances: a more uniform distribution is then rewarded by Eq.~\eqref{eq:EOFss} with a higher electroosmotic flow. On the contrary, at large distances, the SC density \eqref{eq:SCdensity} decays to zero faster than the MF one \eqref{eq:densityPB}: this corresponds to a constant mid-plane velocity for SC (no ion is present to supply a velocity gradient within distances $>\mu$ from the walls) and, conversely, to a virtually diverging velocity far from the charged surfaces for MF (due to the aforementioned algebraic tail). 

It is also worth noticing how the addition of a new length scale, $a$, through the expression of the effective field \eqref{eq:kappach1} affects $\utilde$: the system goes from a velocity \eqref{eq:EOFPB} or \eqref{eq:EOFPBasymptotic}, where the relevant length scale is $\mu$, to a velocity \eqref{eq:EOFsc} or \eqref{eq:EOFscAsymptotic}, where the relevant length scales are $a$ and the geometrical average $\sqrt{\mu a}\propto\mu\Xi^{1/4}$. Indeed, in a strongly correlated system, $\mu\Xi^{1/4}$ ($\gg \mu$) is the length scale over which ions transition from an entropy-favored state with uniform distribution along $z$, to a state where half of the ions are adsorbed on either wall~\cite{Varenna}. Note that upon increasing the coupling parameter $\Xi$, the length scale $a$ can feature different behaviours, depending on which parameter is modified: it may increase if only $q$ is modified, it may decrease if 
correlations are enhanced through an increase of the surface charge, or it may stay constant if temperature is modified. The latter $T$-based view provides a convenient way to envision the role of correlations (decreasing $T$ increases $\Xi$ at fixed geometrical length $a$), but one should keep in mind that it is
experimentally almost impossible to change $T$ significantly, at least with a water solvent.

We now look at the position of the peaks in Fig.~\ref{fig:plotbump}. The spacing $d_\mathrm{max}$ at which $\umax$ is the largest, for a given $\Xi$, can be computed by imposing that the derivative of $\umax$, from Eq.~\eqref{eq:EOFsc}, with respect to $\dtilde=d/\mu$, vanish. This amounts to solving the equation
\begin{equation}
\dtilde^4 + 2\dtilde^2 \Xi - 2\Xi\sqrt{\dtilde^2+\Xi}\, \sinh\left({\frac{\dtilde^2}{2\sqrt{\dtilde^2+\Xi}}}\right) =0 \,.
\end{equation}
Making the assumption that $\dtilde^2\ll\Xi$, for $\Xi$ sufficiently large, this reduces to
\begin{equation}
\sinh\left({\frac{\dtilde^2}{2\sqrt{\Xi}}}\right) \simeq \frac{\dtilde^2}{\sqrt{\Xi}}\,.
\end{equation}
The equation $\sinh(y/2)=y$ has only one real positive solution: $y_0=4.3546...\,$. By equating $\dtilde^2/\sqrt{\Xi}$ to $y_0$, one gets
\begin{equation}
\label{eq:dmax}
\dtilde_{\mathrm{max}}\simeq \sqrt{4.3546}\,\, \Xi^{1/4}\,,
\end{equation} 
which satisfies the underlying assumption $\dtilde^2\ll\Xi$ for large $\Xi$ and is consistent with Eq.~\eqref{eq:EOFscAsymptotic}. 
Eq.~\eqref{eq:dmax} agrees with numerical calculations of $\dtilde_\mathrm{max}$ from Eq.~\eqref{eq:EOFsc}.
Eqs.~\eqref{eq:EOFscAsymptotic} and \eqref{eq:dmax} together indicate that the maximum attainable velocity $\utilde_\mathrm{m,max}$, i.e.~$\umax$ at $d_{\mathrm{max}}$, also scales as $\Xi^{1/4}$. Compared to Eq.~\eqref{eq:EOFPBasymptotic}, this indicates enhanced flow with respect to the MF prediction.

\begin{figure}[]
\centering
\includegraphics[width=\columnwidth]{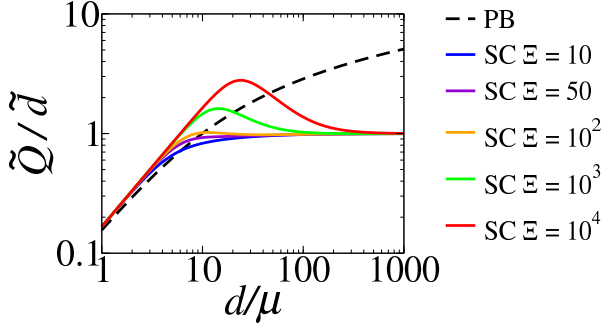}
\caption{
Integrated flow divided by distance, as a function of $d/\mu=\dtilde $, in log-log scale. The situation is analogous to that of Fig.~\ref{fig:plotbump}. 
}
\label{fig:plotQ}
\end{figure}

A $d_\mathrm{max}$ scaling as $\mu\Xi^{1/4}$ confirms our previous interpretation. Increasing the inter-plate distance when ions are uniformly distributed, i.e.~when $d<\mu\Xi^{1/4}$, increases linearly the electroosmotic speed at the midplane: indeed, the uniform ion density scales as $1/d$, so does the curvature of the electroosmotic profile, and therefore $\umax= \utilde(d/2)$ grows as $d$. Increasing the distance beyond $\mu\Xi^{1/4}$, though, destroys uniformity and gradually favors an ionic distribution decaying rapidly away from the walls: at constant total applied force ($\int_0^d n(z)\,\mathrm{d}z$ is fixed by electroneutrality), this configuration gives a lower $\umax$ than the uniform distribution configuration.

This non-monotonic effect of correlations is visible in integrated quantities, too. To see this, we compute the volume flow rate $Q$, as we did for the MF case. By integrating Eq.~\eqref{eq:EOFscz} for $z$ between $0$ and $d$, we get
\begin{align}
\Qtilde\,=&\,-\frac{2 \Xi }{\dtilde^2}+\sqrt{\dtilde^2+\Xi } \coth \left(\frac{\dtilde^2}{2 \sqrt{\dtilde^2+\Xi}}\right)-2 \nonumber \\ 
=\,&\,
\begin{dcases}
\frac{d^2}{6 \mu^2} & \quad \text{if}\ d\ll\mu\Xi^{1/4} \\
\frac{d}{\mu} & \quad \text{if}\ d\gtrsim \mu\Xi^{1/2}
\end{dcases}\,,
\label{eq:Q}
\end{align}
where we used the same units as in Eq.~\eqref{eq:Qunit}.
Fig.~\ref{fig:plotQ} represents $\Qtilde$ for different values of $\Xi$, and for MF (Eq.~\eqref{eq:QPB}). 

\begin{figure}
\includegraphics[width=\columnwidth]{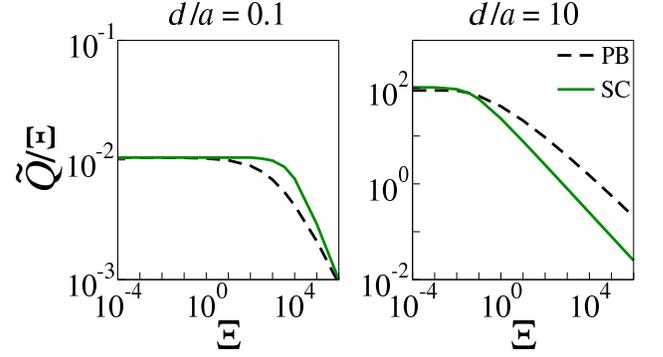}
\caption{Integrated flow $Q\,2\pi\eta /(q e \Epar) = \Qtilde/\Xi$, as a function of coupling.
Two values of separation $d$ are shown. In both cases, as correlations are switched on, the SC flow (accurate for large $\Xi$) decreases compared to the correlation-free value (PB curve, $\Xi\to 0$). However, at small separation ($d/a=0.1$, left), 
the SC theory predicts a stronger flow than PB. At large separations ($d/a=10$, right), this is not true and the effect shown in the large-$d$ part of Fig.~\ref{fig:plotQ} dominates. }
\label{fig:fluxalt}
\end{figure}

\begin{figure}
\includegraphics[width=\columnwidth]{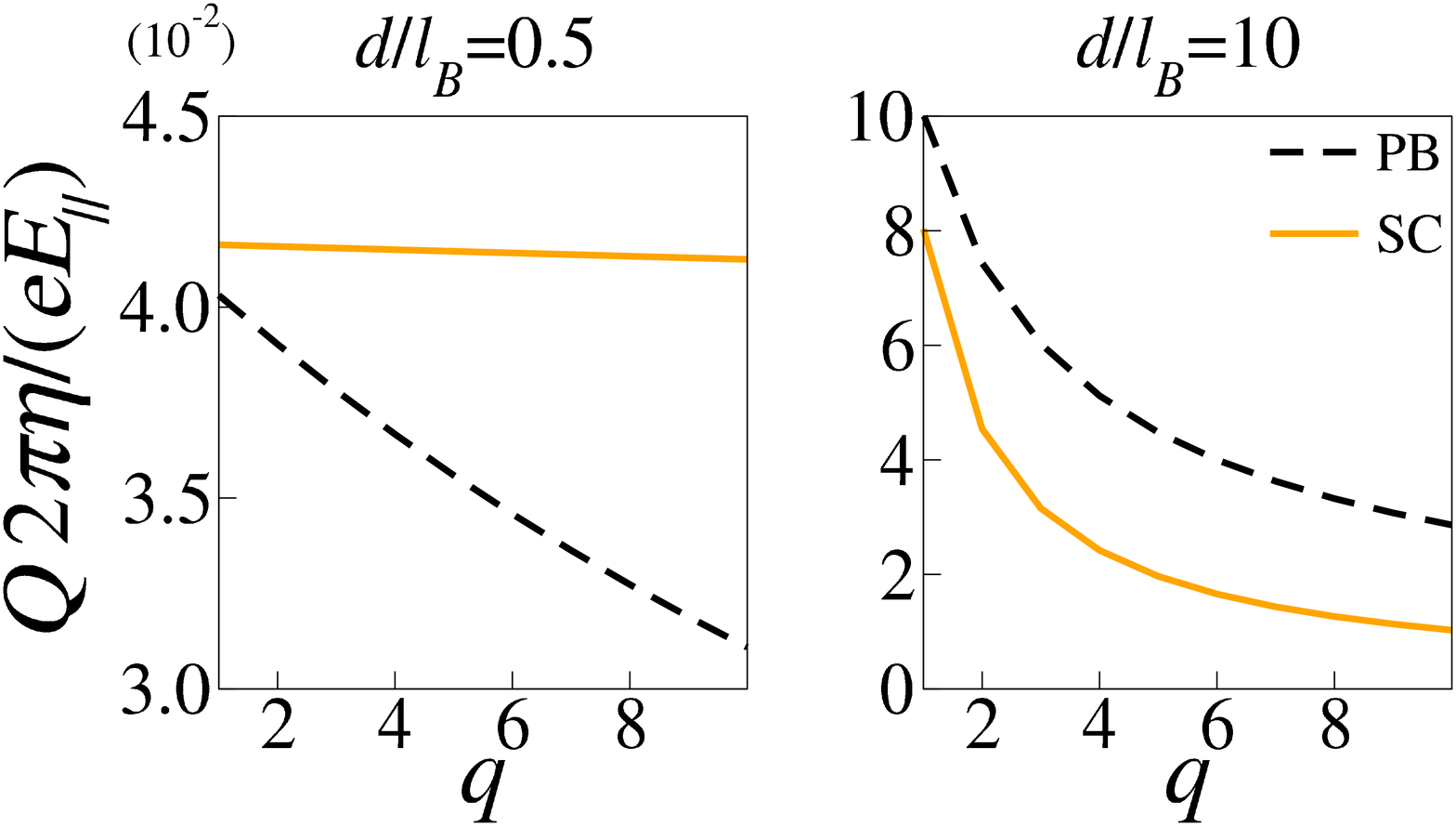}
\caption{Integrated flow $Q\,2\pi\eta /(e \Epar)$, as a function of valence $q$. Here, $2\pi\lB^2\sigma=1$ is kept fixed, so that the horizontal axis can also be read as $\Xi^{1/3}$. Two values of separation $d$ are shown.}
\label{fig:fluxalt2}
\end{figure}

To better understand the effect of correlations, we look at $Q$ in real units, that do not involve $\lB$. We fix the separation $d$, in units of the purely geometrical parameter $a$, and imagine to gradually increase coupling by increasing $\lB$ only. 
As mentioned above, this provides us with a theoretical tool to gradually switch on correlations at constant separation and assess their effect.  
Fig.~\ref{fig:fluxalt} shows that a correlated system exhibits in general a smaller flow than the uncorrelated one. This is also evident from the fact that the maximum attainable velocity $\utilde_\mathrm{m,max}$ scales as $\Xi^{1/4}$; when phrased in terms of the original unscaled velocity from definition \eqref{eq:uunit}, this translates to $u_\mathrm{m,max}\propto \lB^{-1/2}$: increasing correlation effects at constant charge thus leads to a net decrease of transport. 
However, a correlated system, well described by the SC theory, exhibits at small separations a stronger flow than predicted by applying MF theory to the same system (Fig.~\ref{fig:fluxalt}, left). In this sense and in this regime, correlations enhance electroosmosis.

Finally, yet another way to assess the role of correlations is to observe how the flux changes as a function of ion valence $q$, keeping constant $\lB$ and $\sigma$. This amounts to increasing $\Xi$ only through $q$. Fig.~\ref{fig:fluxalt2} shows the same qualitative behaviour as Fig.~\ref{fig:fluxalt}: it exhibits at small separation $d$ an enhancement and at large separation a suppression of the electroosmotic flow, compared to the MF prediction.

\section{Computational model}
\label{sec:Sim}

\subsection{Simulation setup}

\begin{figure}[]
\includegraphics[width=\columnwidth]{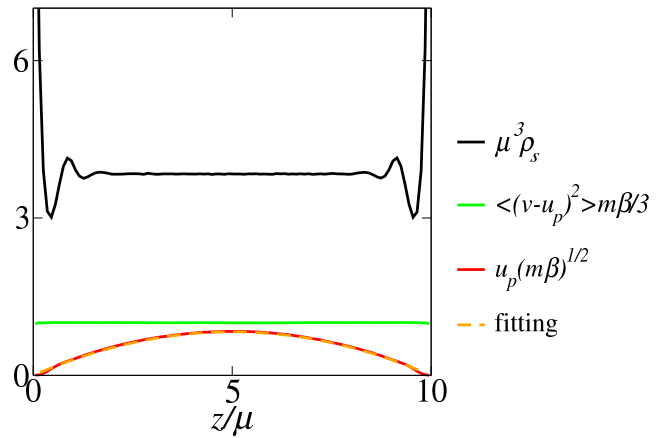}
\caption{Example of pure DPD solvent particles simulations (uncharged system). The concentration and temperature profiles are represented by solid lines. Temperature is measured locally from the particles' velocity variance. 
Symbols are the parallel velocity profile $u_p$, while the dashed line is the fitting curve. The parameters are $\mu^3 n_s=4$, $\beta\mu F_\parallel=0.02$ and $d/\mu=10$. Depending on $\Xi$, different parameters are used in order to reduce the total number of particles.}
\label{fig2}
\end{figure}

The DPD simulations are performed in a box of size $L_{xy} \times L_{xy} \times L_z$ with periodic boundary conditions in the $x$ and $y$ directions. The charged planes with charge density $\sigma$ are located at $z=0$ and $z=d$, as above. $N_c$ counterions and $N_s$ DPD particles (modeling solvent) are confined in region $0<z<d$, with $d\leq L_z/2$ and $L_z\geq L_{xy}/2$. The vacuum region in the $z$ direction is defined according to a recently developed Ewald sum method~\cite{DoGi16}. The $N_s$ solvent particles correspond to a solvent number density $n_s$. The $N_c$ counterions are considered as DPD particles with a centered charge $qe$. The parameters and method of integration used in our DPD simulations, widely used in the literature, are given in~\cite{GrWa97}. The same mass $m$ is used for ions and DPD solvent particles. Besides $m$, the thermal energy $\beta=1/k_BT$ and the Gouy-Chapman length $\mu$ are the natural units in our calculations.
Considering the level of coarse graining in our method and the focus on hydrodynamic phenomena, important properties of water in confinement such as dielectric response and water molecules alignment~\cite{Schlaich2019} are not taken into account.

The dynamic viscosity of the DPD system is determined from the fitting of the velocity profile $u_p(z)$ of a pure system (solvent particles only), in a Poiseuille-Hagen flow configuration:
constant and uniform flow force $\Fpar$, parallel to the surfaces, under no-slip boundary conditions. The velocity profile parallel to the surfaces then reads
\begin{equation}
u_p(z)=\gamma\left[\frac{d^2}{4} - \left(z-\frac{d}{2}\right)^2\right] \ .
\end{equation}
The viscosity expression is well studied in~\cite{SmAl08,SmSe09}. It is given by
\begin{equation}
\eta=\frac{n_s \Fpar}{2\gamma} \ .
\end{equation}
In Fig.~\ref{fig2} we show the results for one of the pure systems considered. The dimensionless particle concentration is $\mu^3 \rho_s$ and the temperature profile is given by $\langle v^2\rangle m\beta$, where $\langle v^2\rangle$ is the average quadratic velocity. The parallel velocity $u_p(z)$ and the fitting curve are also shown. Following the fitting procedure, we find $\gamma=0.033~(m\beta)^{-1/2}\mu^{-2}$ and $\eta=1.21~(m/\beta)^{1/2}\mu^{-2}$.

In order to induce the electroosmotic flow in DPD simulations with counterions, the dynamics is performed with a constant electric field $\Epar=1\ (\beta \mu q e)^{-1}$ parallel to the surfaces, as in Fig.~\ref{fig1}. The lateral side size of the simulation box is chosen as $L_{xy}/\mu=\sqrt{\pi\Xi N_c}$, so as to maintain charge neutrality.
For higher coupling parameters $\Xi$, in order to keep charge neutrality, the size of the simulation box can be very high. This means that a typical concentration of DPD particles, such as $\mu^3 n_s=4$, would give a huge number of particles. To circumvent this issue, we decrease the concentration (until we get a computationally reasonable number of particles, around $4\times 10^4$) and increase the DPD interaction cut-off radius to obtain the desired hydrodynamic properties. We concomitantly increase the level of coarse graining. As a consequence we must obtain the viscosity for each set of parameters.
No-slip boundary conditions are applied. The electrostatic part of molecular dynamics is also derived from the modified Ewald sum method in~\cite{DoGi16}. The fraction of number of ions/solvent particles in simulations goes approximately  from $0.01$ to $0.001$, depending on the coupling parameter.

\subsection{Results}

\begin{figure*}[]
\begin{center}
\includegraphics[scale=0.2]{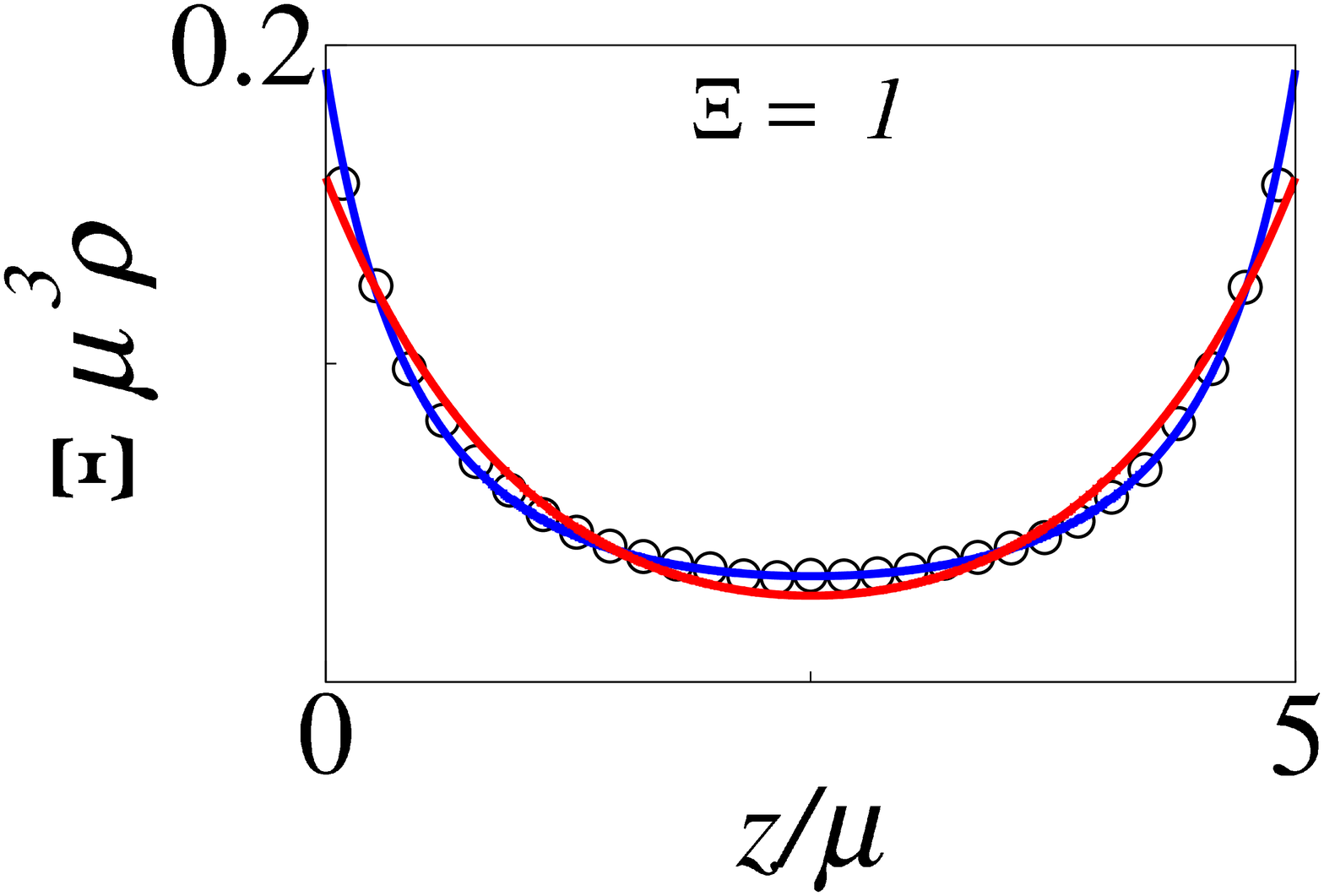}
\includegraphics[scale=0.2]{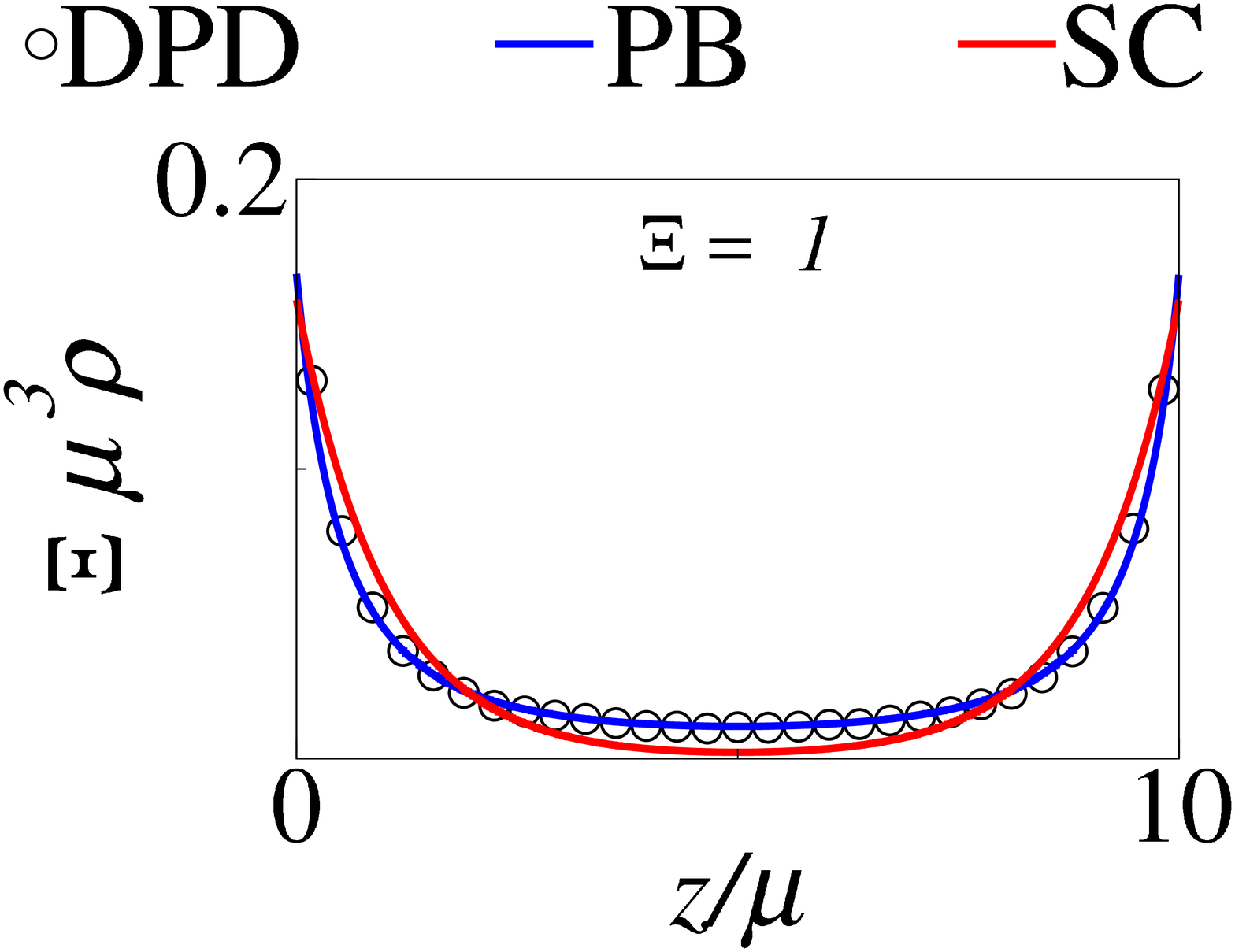}
\includegraphics[scale=0.2]{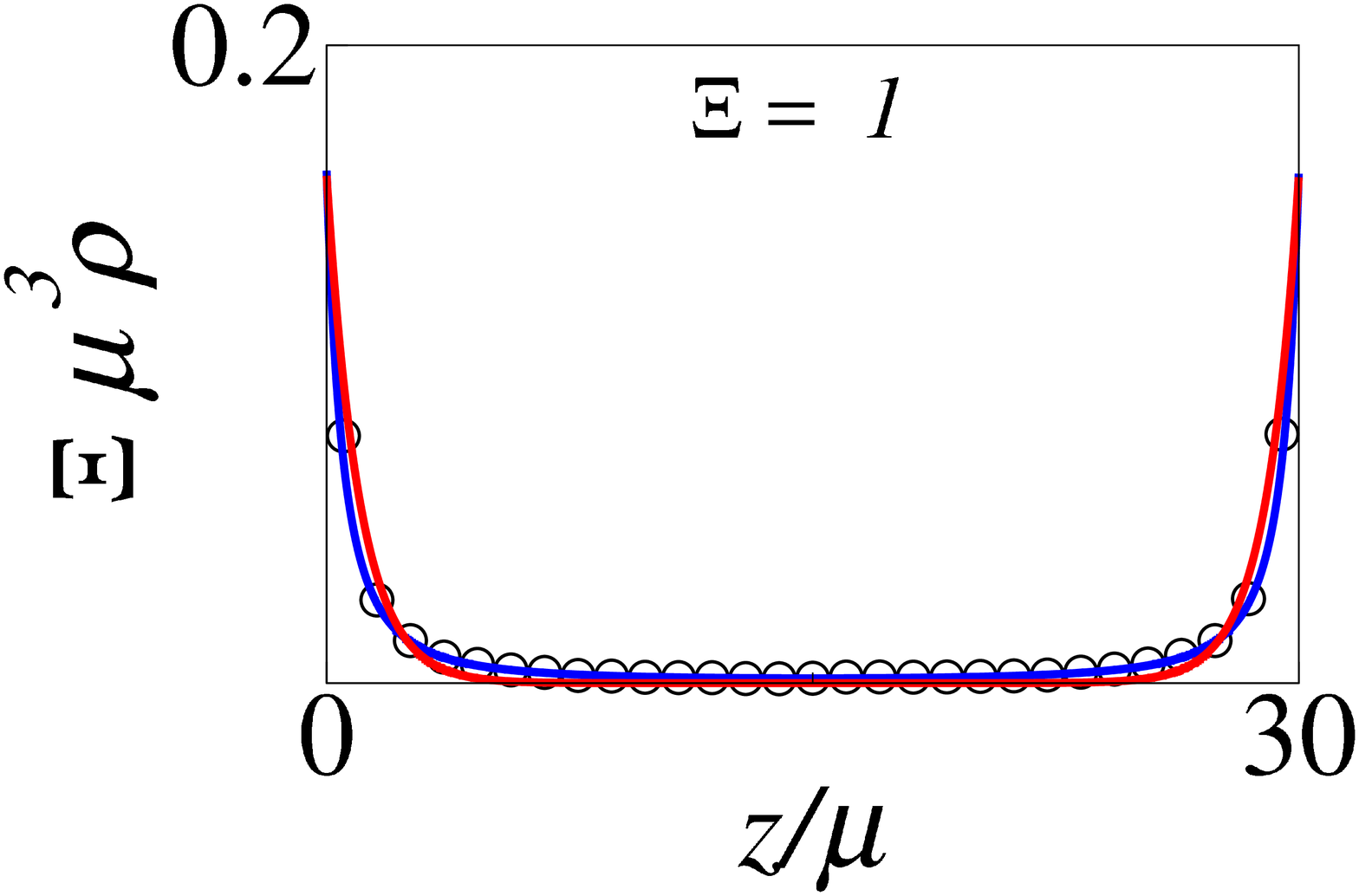}

\includegraphics[scale=0.2]{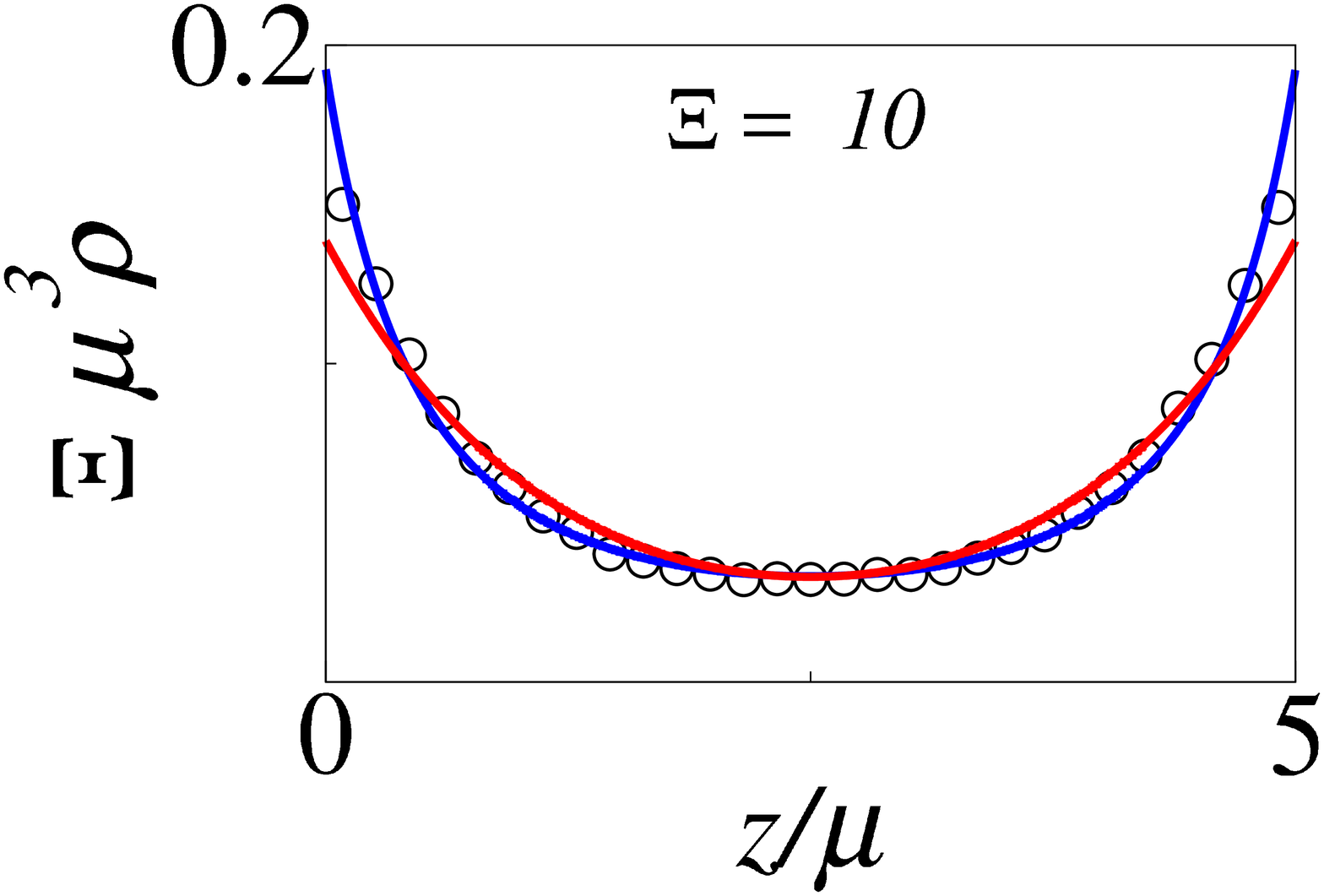}
\includegraphics[scale=0.2]{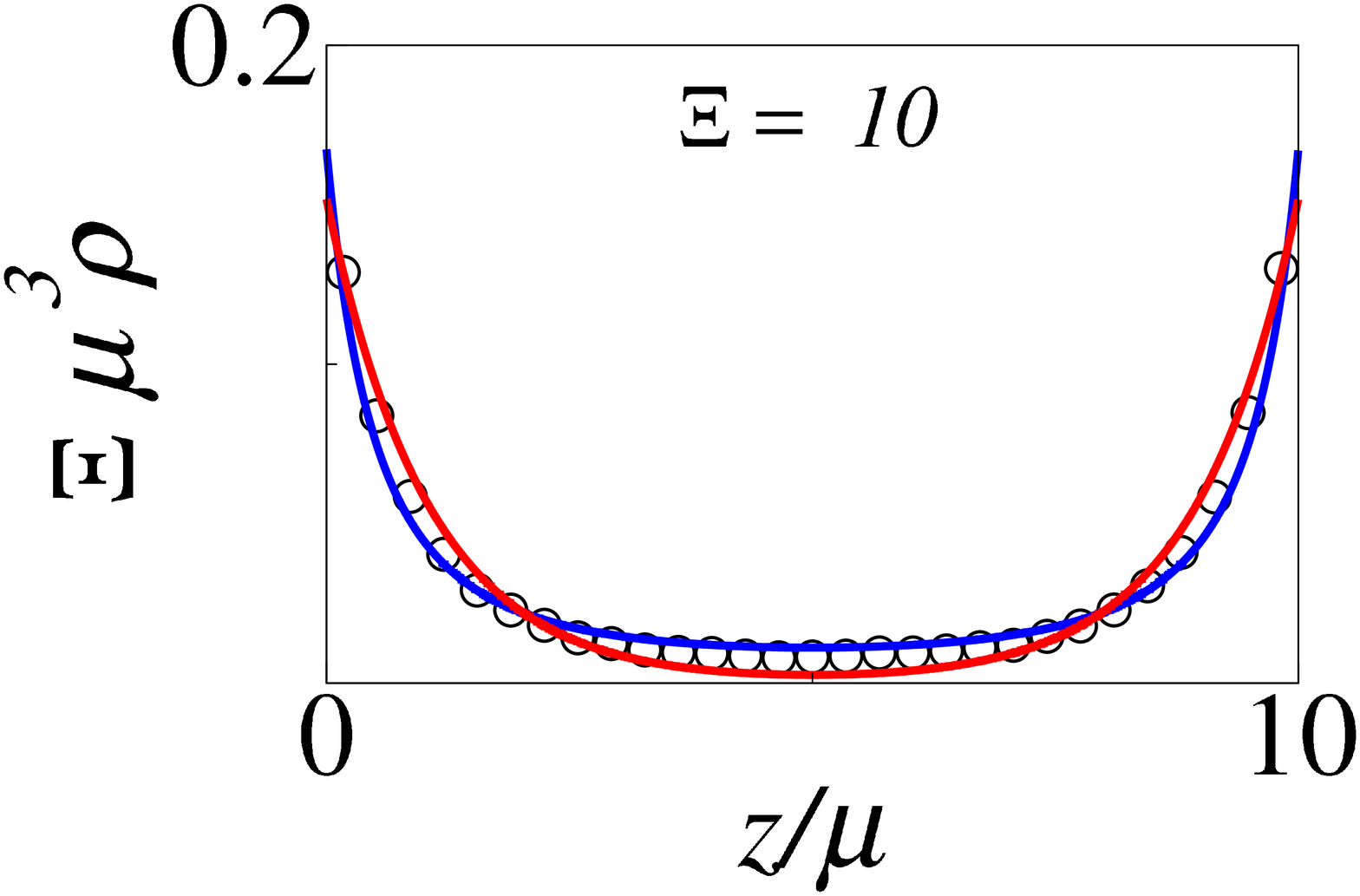}
\includegraphics[scale=0.2]{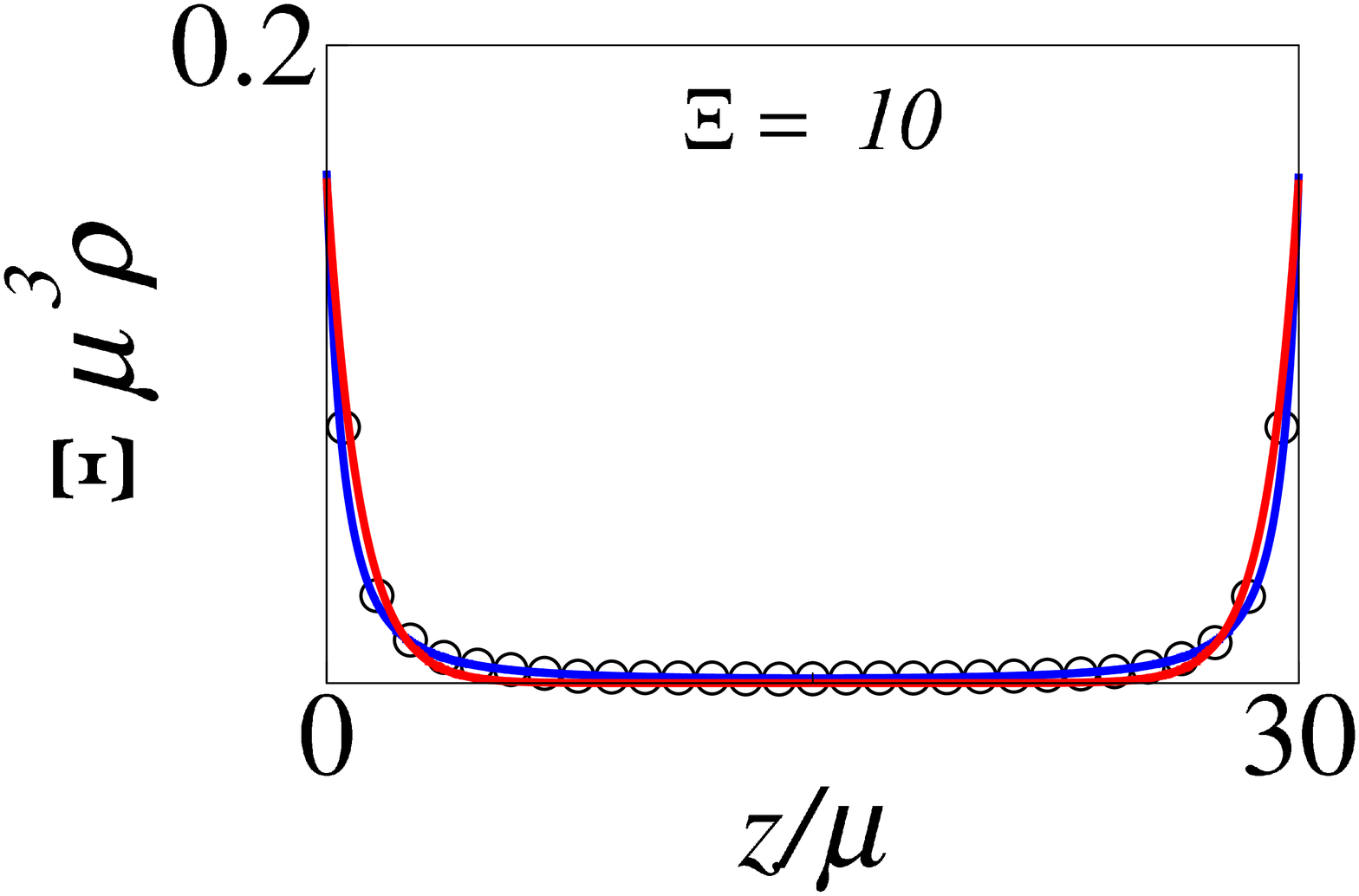}

\includegraphics[scale=0.2]{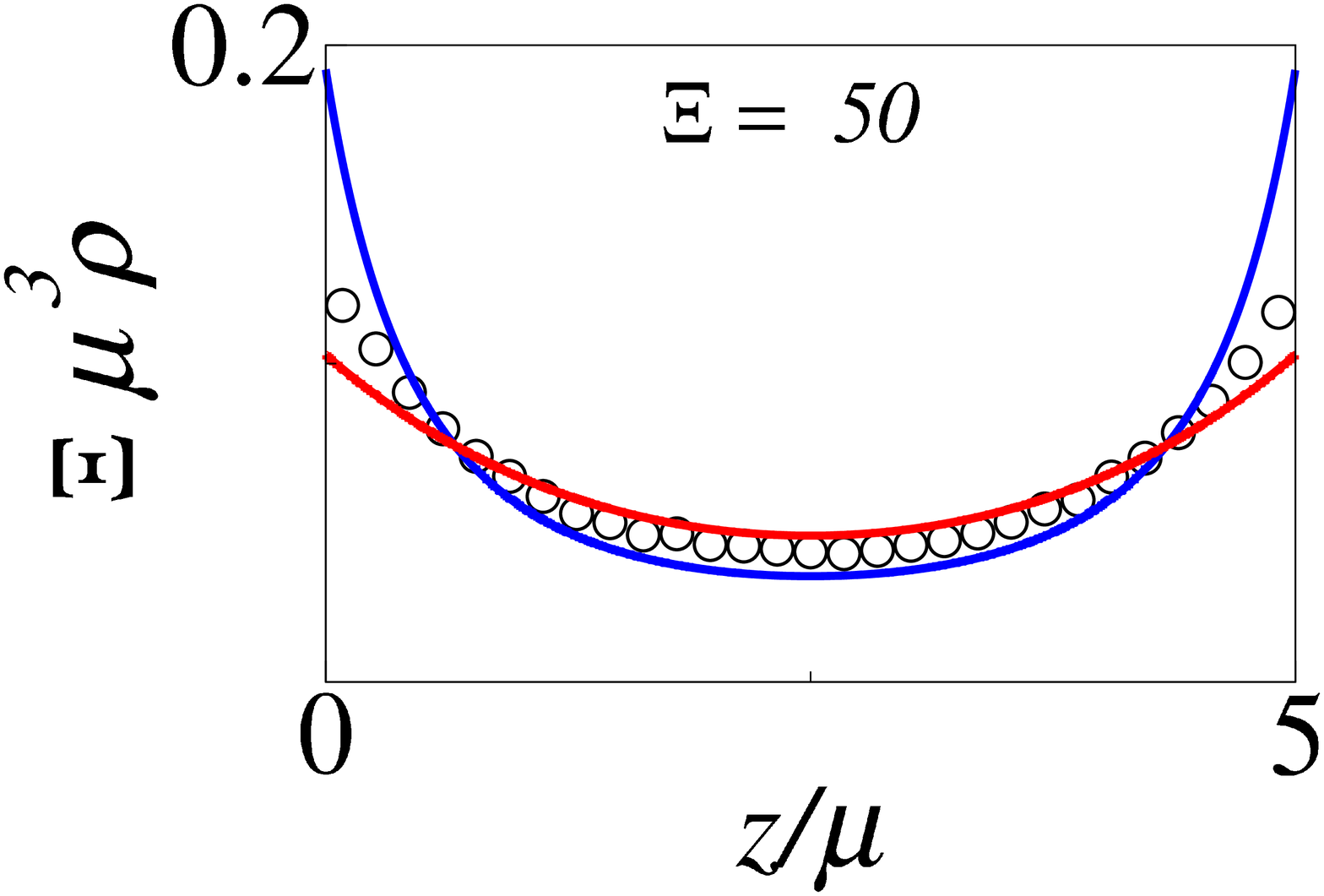}
\includegraphics[scale=0.2]{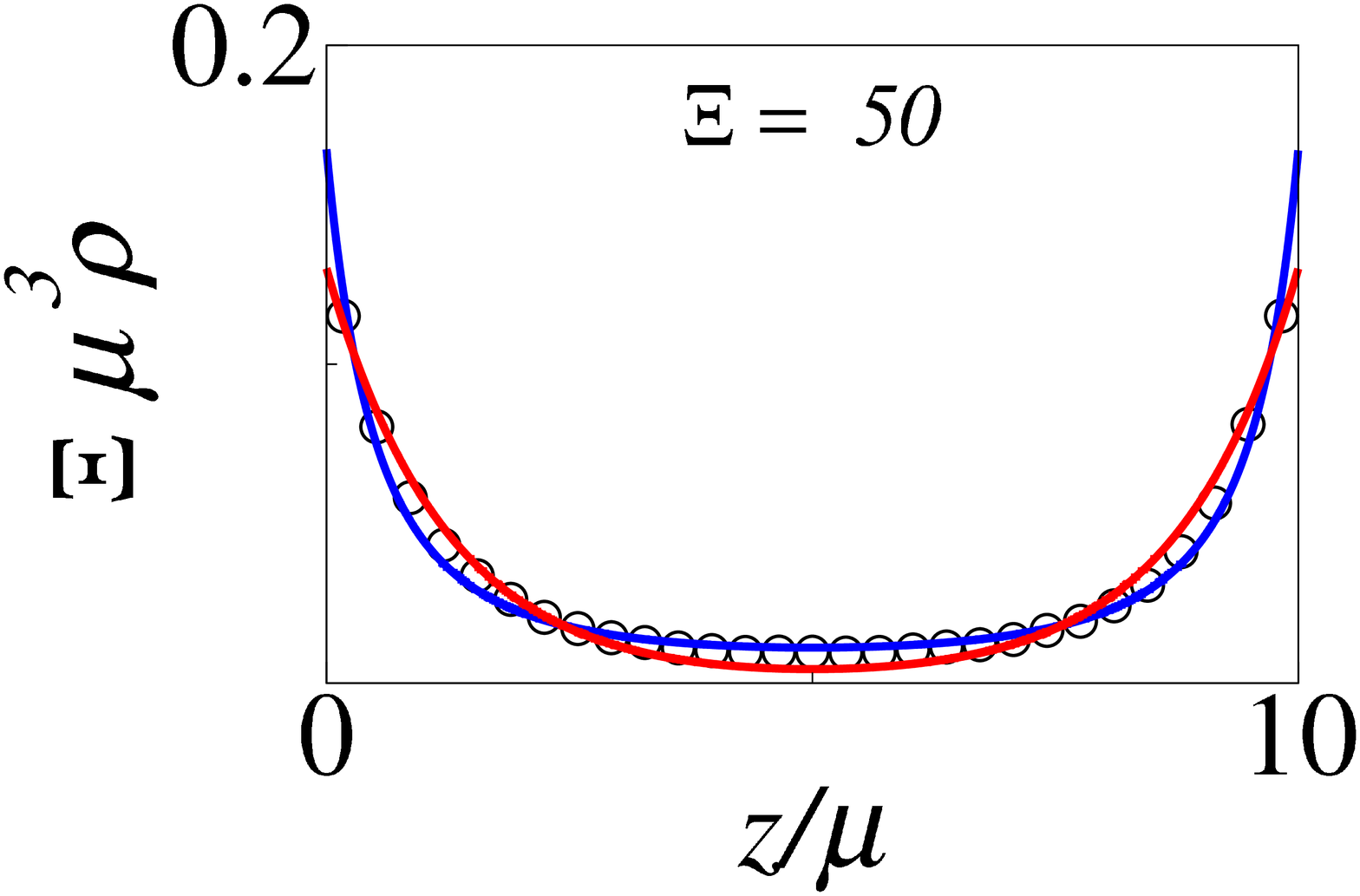}
\includegraphics[scale=0.2]{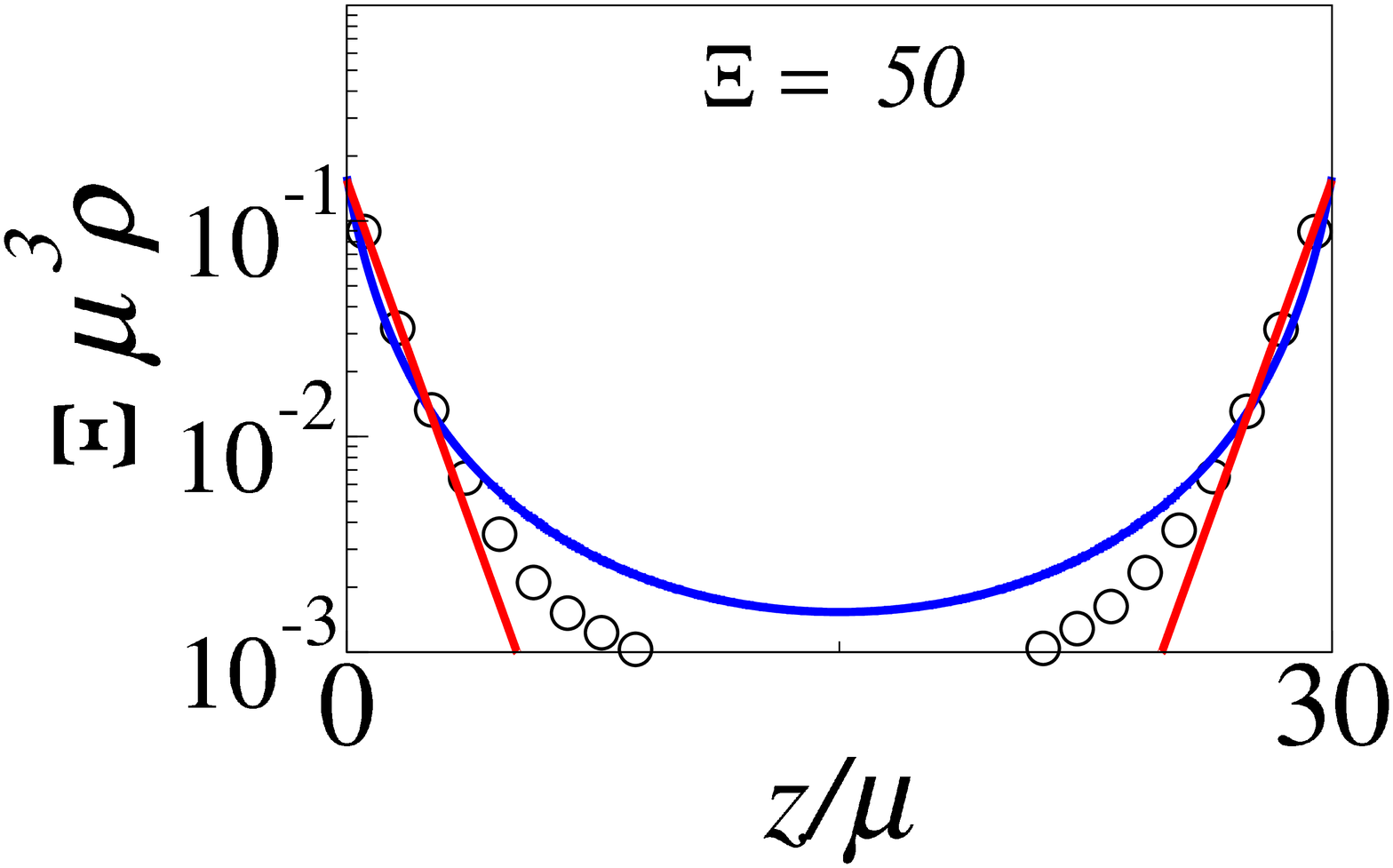}

\includegraphics[scale=0.2]{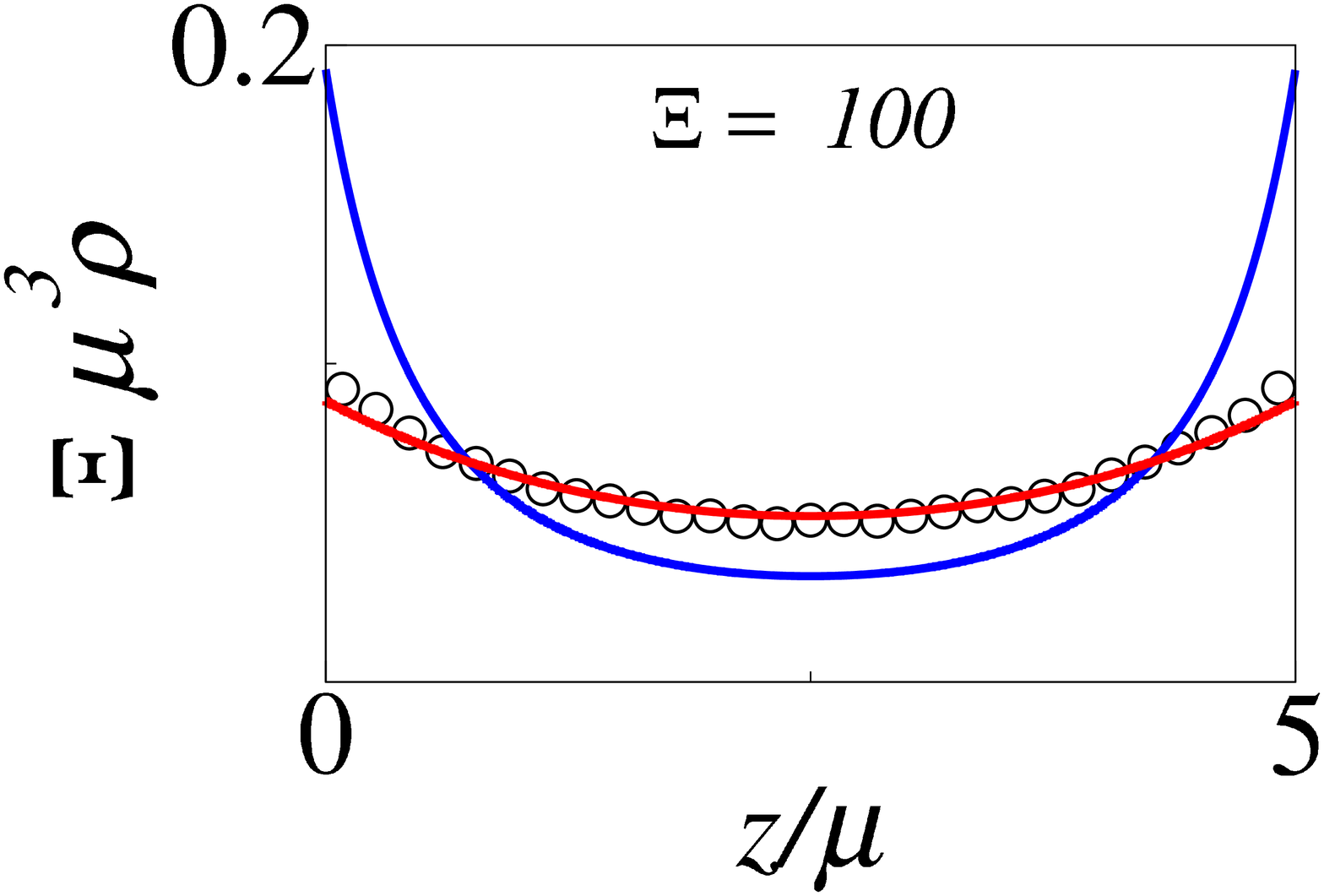}
\includegraphics[scale=0.2]{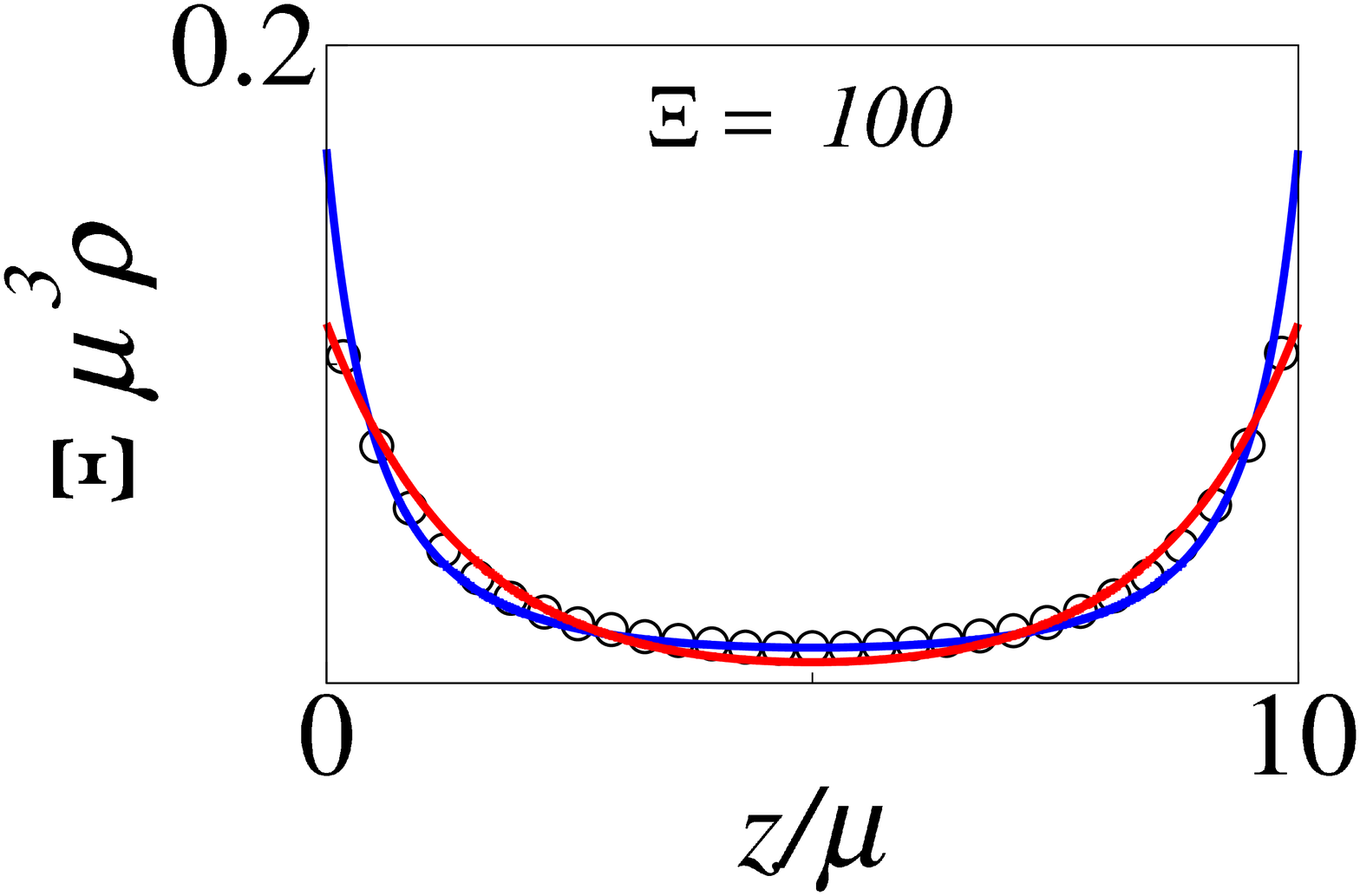}
\includegraphics[scale=0.2]{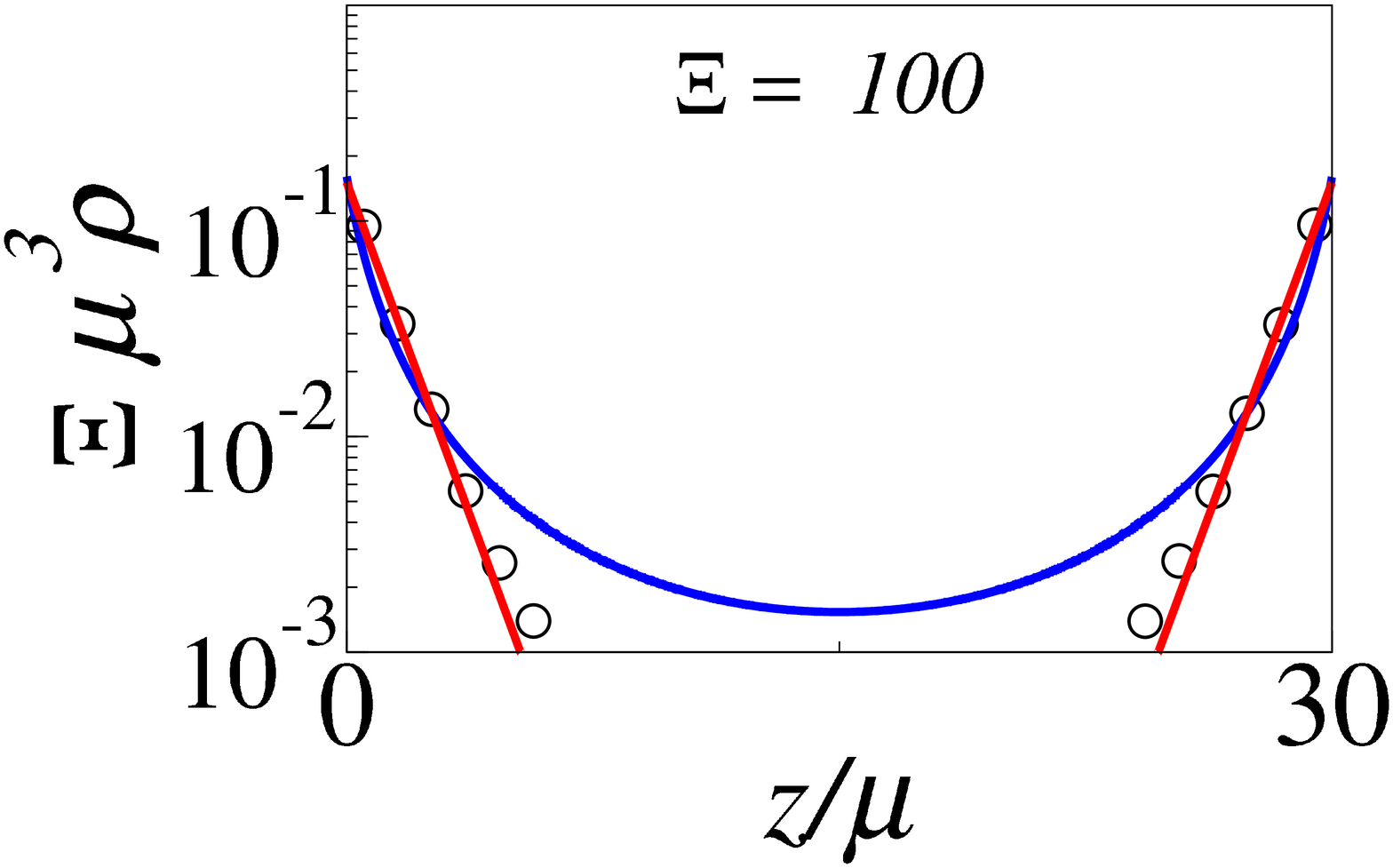}

\caption{Counterions density profiles for coupling parameters $\Xi=1$, $\Xi=10$, $\Xi=50$ and $\Xi=100$ in first, second, third and fourth lines, respectively. The separations are $d/\mu=5$, $10$ and $30$, first, second and third columns, respectively. Symbols are DPD simulation data, whereas lines are for PB and SC theories.}
\label{fig3}
\end{center}
\end{figure*}
\begin{figure*}[]
\begin{center}
\includegraphics[scale=0.2]{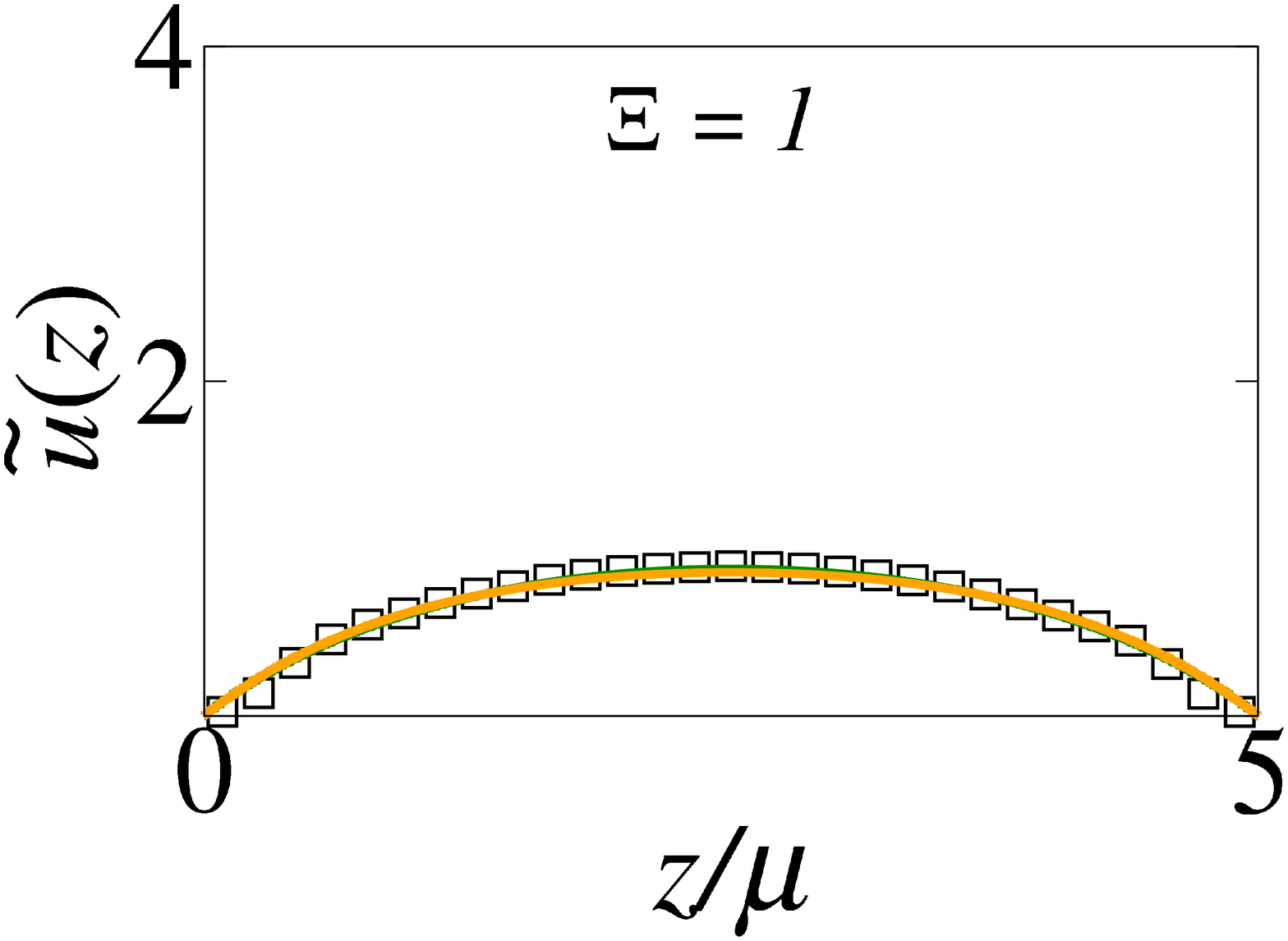}
\includegraphics[scale=0.2]{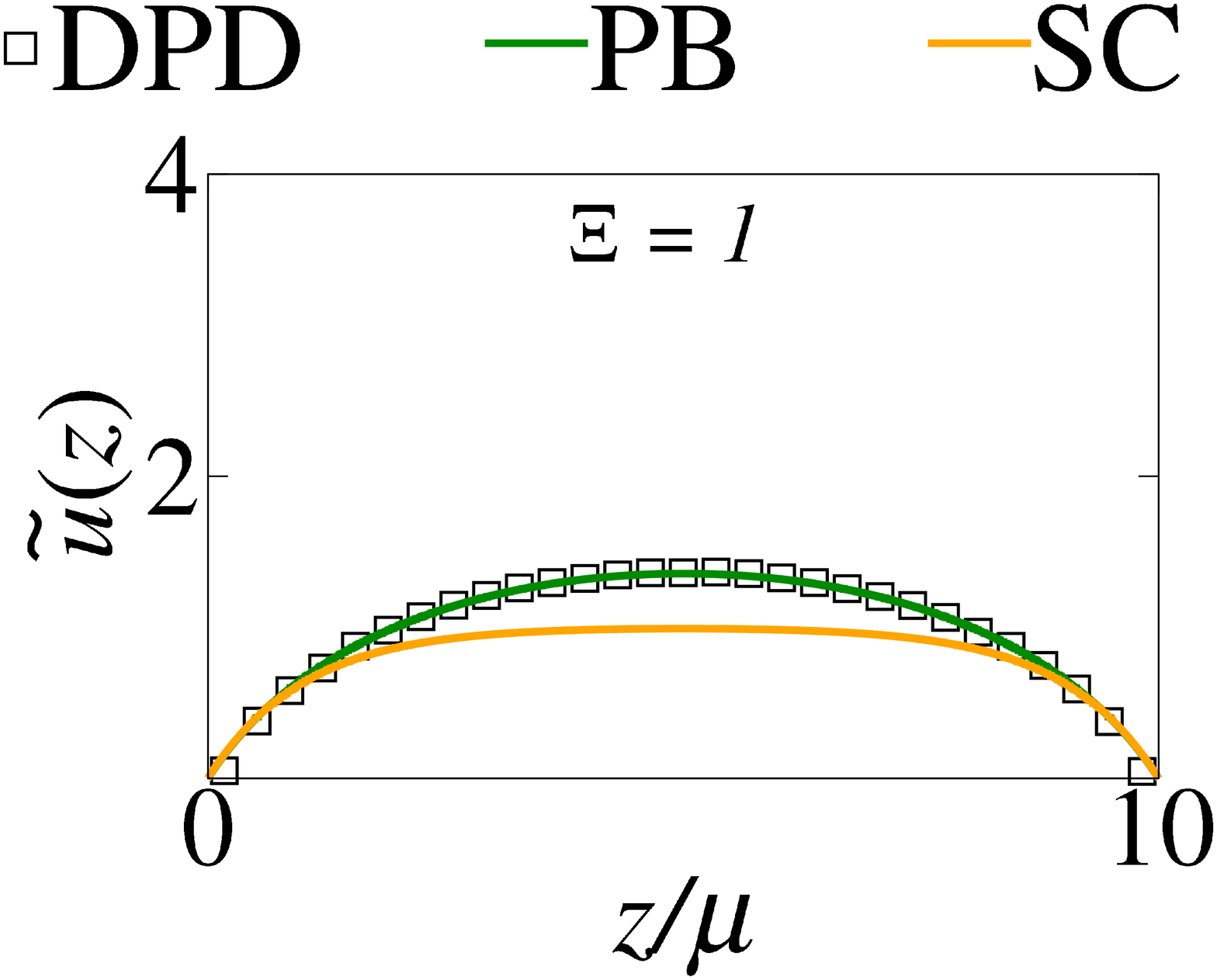}
\includegraphics[scale=0.2]{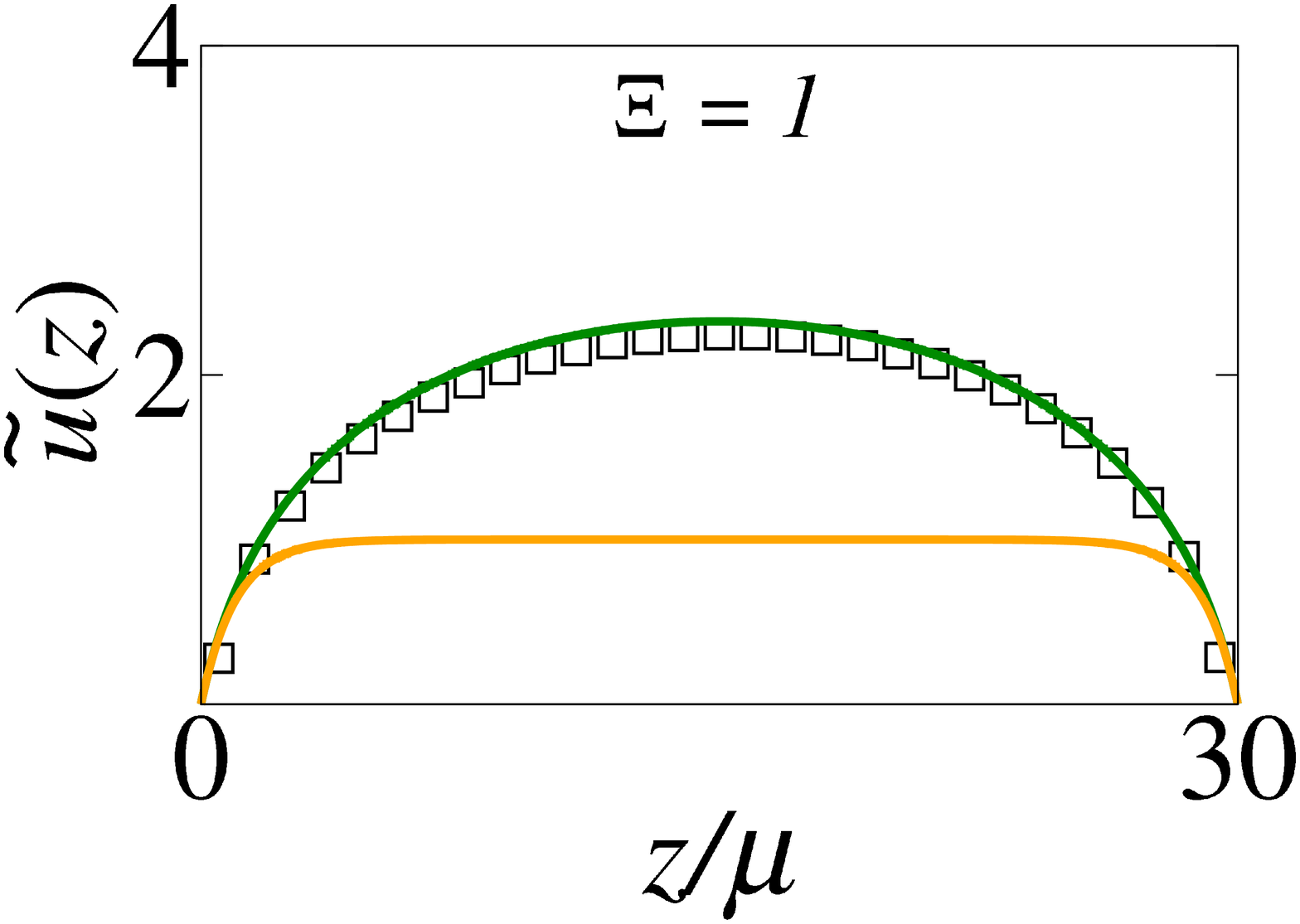}

\includegraphics[scale=0.2]{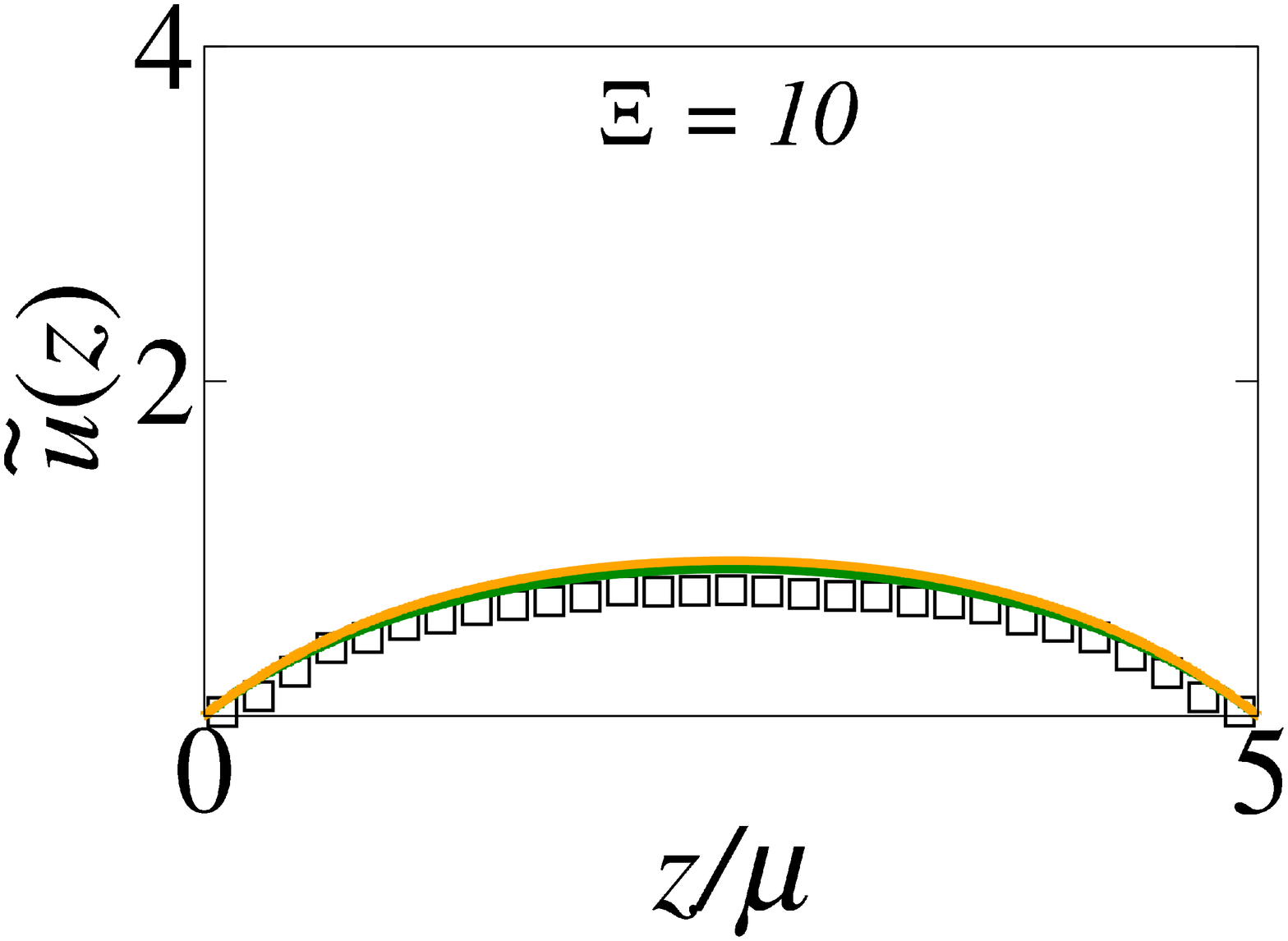}
\includegraphics[scale=0.2]{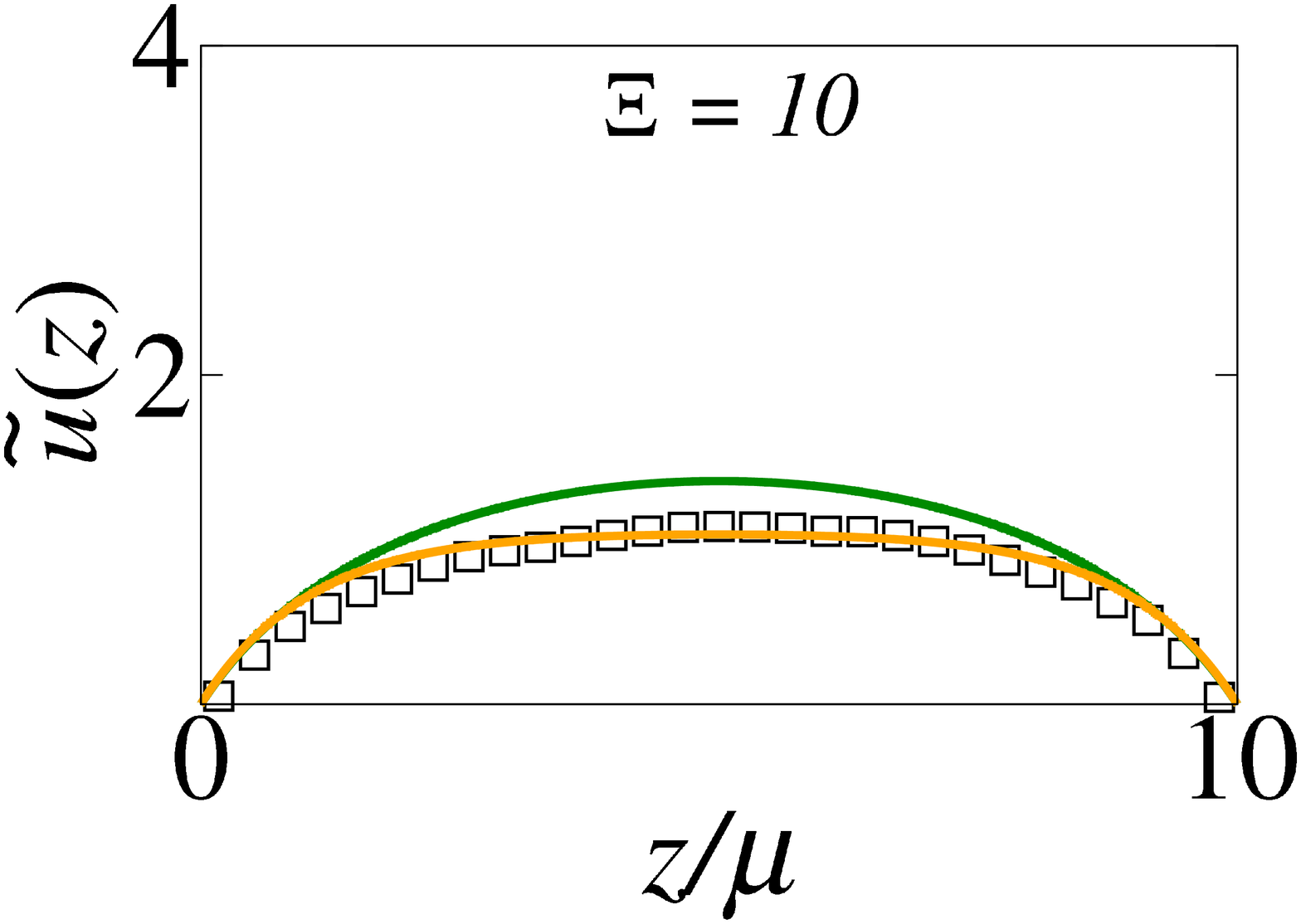}
\includegraphics[scale=0.2]{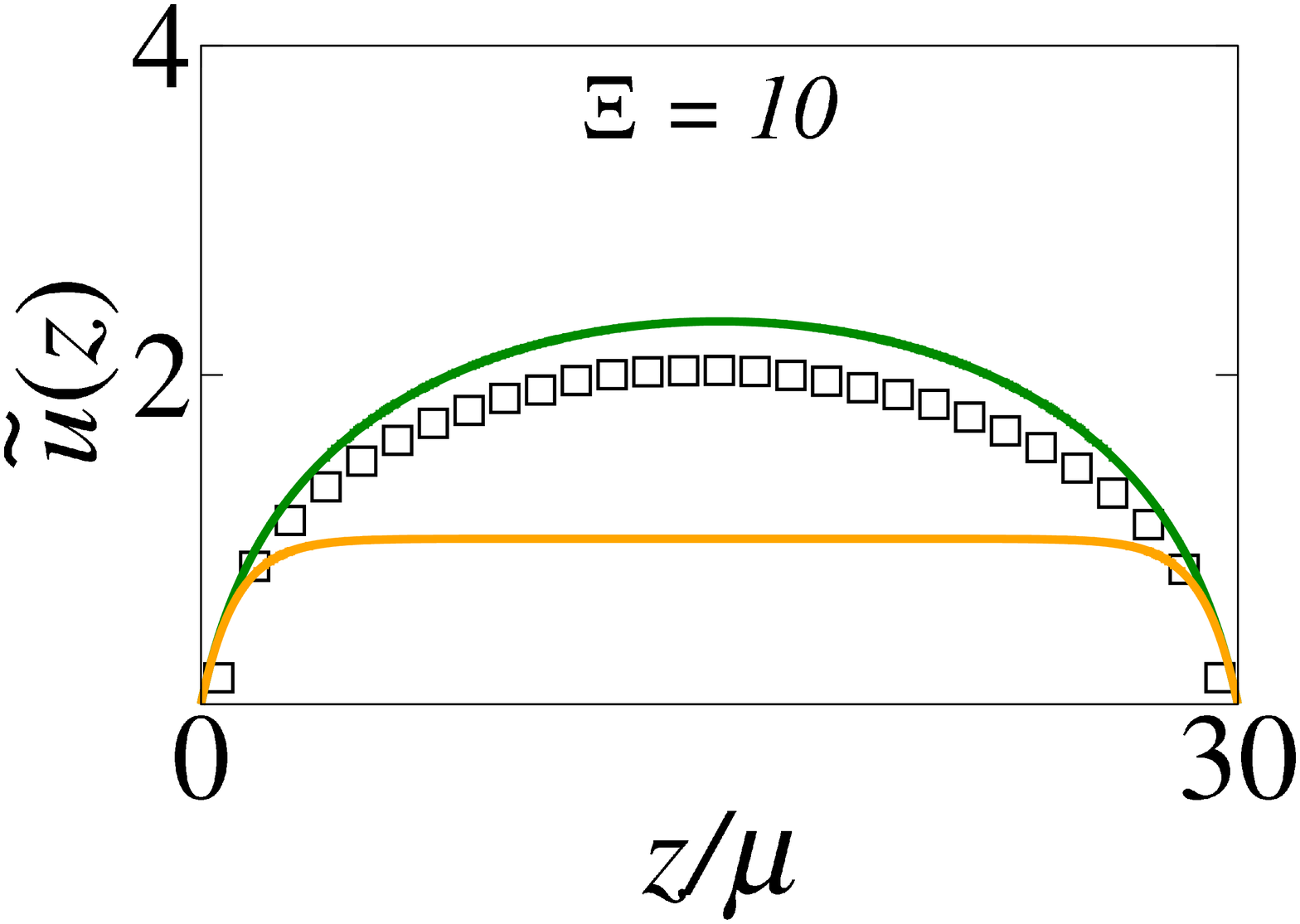}

\includegraphics[scale=0.2]{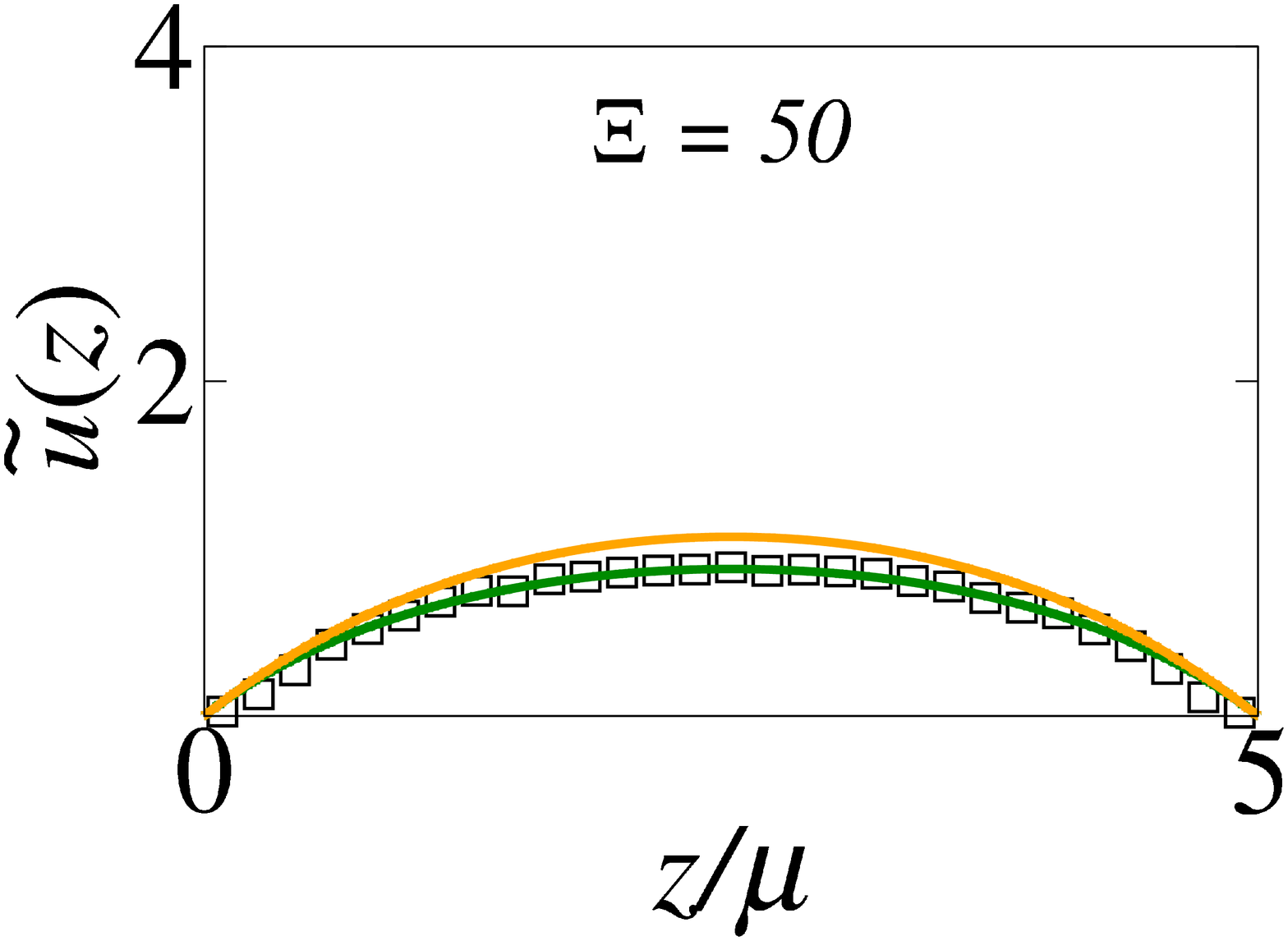}
\includegraphics[scale=0.2]{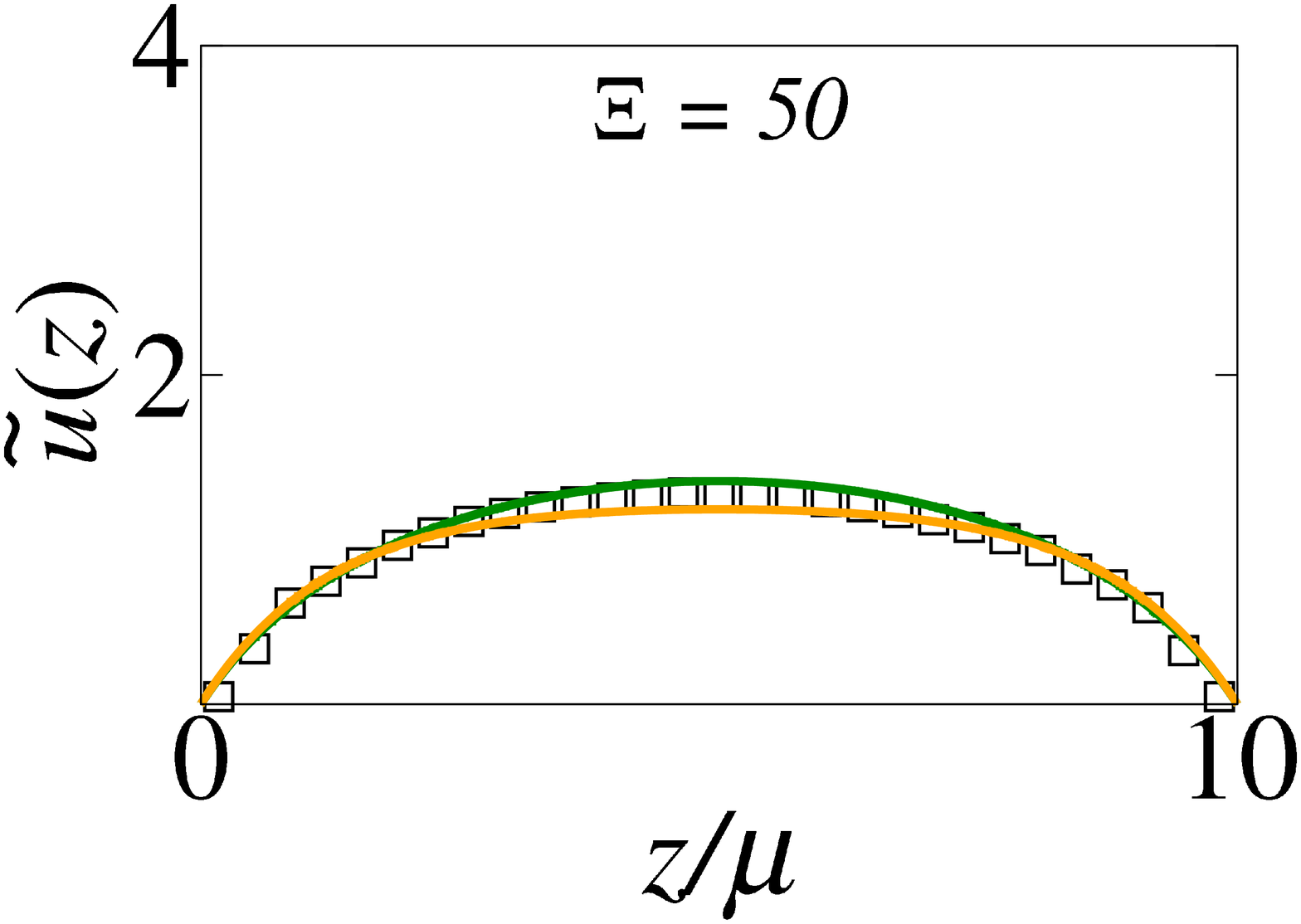}
\includegraphics[scale=0.2]{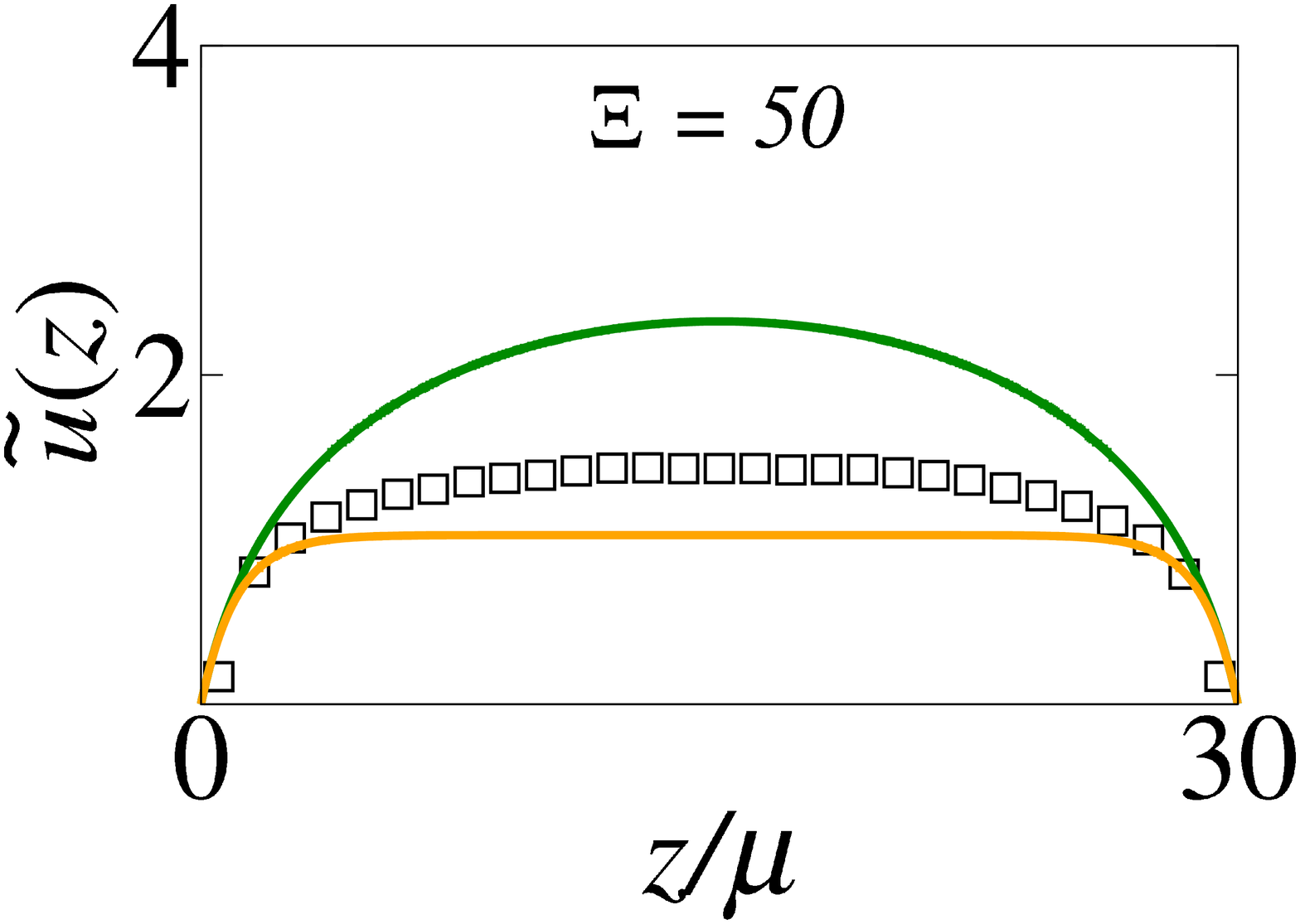}

\includegraphics[scale=0.2]{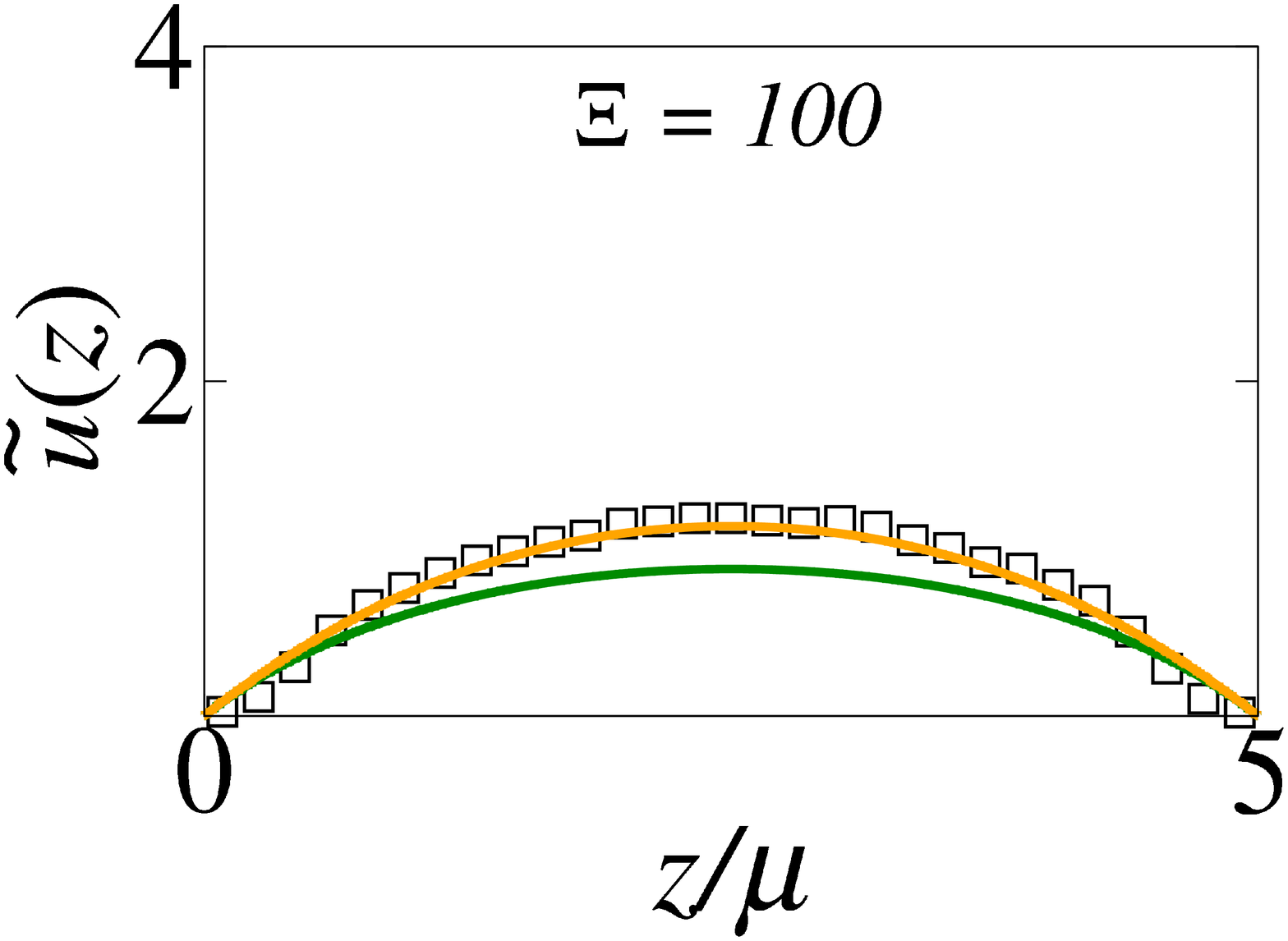}
\includegraphics[scale=0.2]{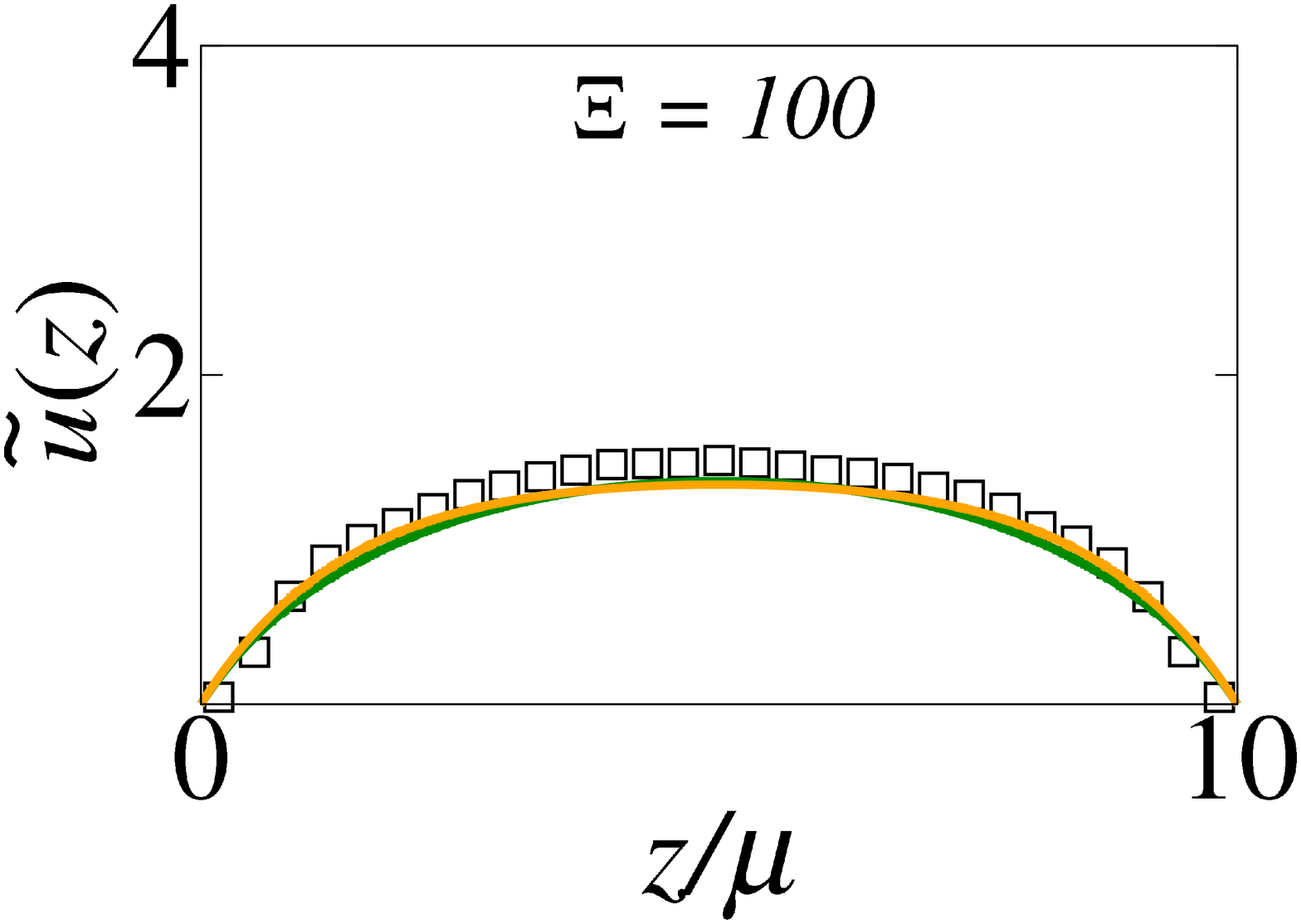}
\includegraphics[scale=0.2]{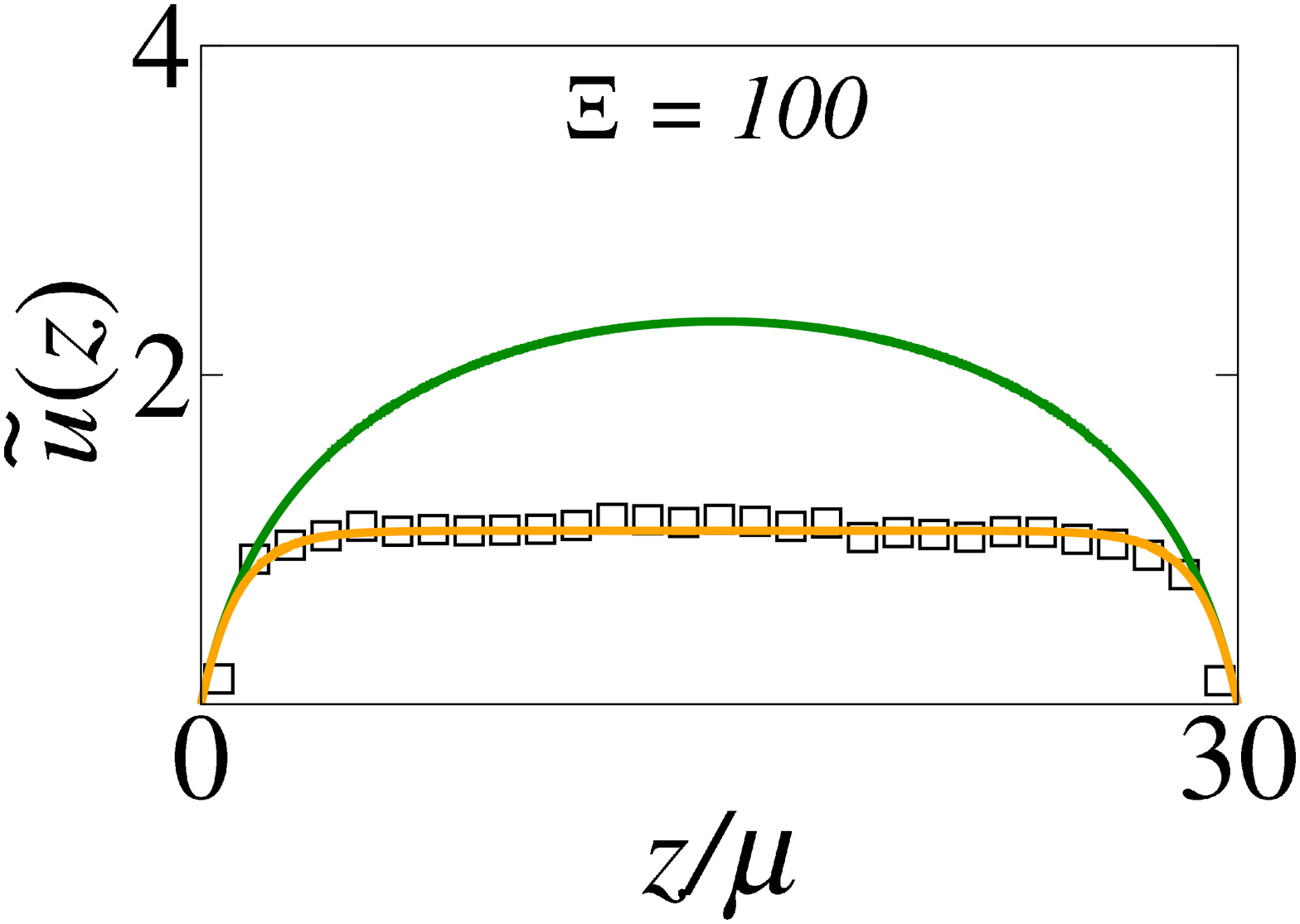}

\caption{Velocity profiles for the same parameters as indicated in Fig.~\ref{fig3}.}
\label{fig4}
\end{center}
\end{figure*}

First, we compare the ionic density profiles obtained from the PB and SC theories with the profiles obtained from the present DPD simulations. Results are represented in Fig.~\ref{fig3}. We sample four values of $\Xi$, ranging from 1 to 100, and three values of $d$, from 5 to $30\mu$. The profiles are not affected by the flow, in line with the discussion in Sec.~\ref{sec:Theory}. For low coupling parameters (first and, to a lesser extent, second line) the PB framework describes our density profiles.  As coupling increases, the profiles depart from the PB result and approach their SC counterpart. 
At $\Xi=100$, 
the effective-field SC theory accurately describes the numerical density profiles. Unfortunately we could not probe larger $\Xi$ values, as this would require to increase substantially the number of particles in simulations. Our results are consistent with the validity limit of the effective field theory, probed with Monte Carlo simulations and better discussed in~\cite{Samaj2018a}; we point out that such limit reaches down to $\Xi$ values that are surprisingly low, given how the theory stems mainly from low-temperature arguments. 

As discussed in Sec.~\ref{sec:Theory}, at small $d$, ions are rather localised on the walls for low values of $\Xi$, while they are almost uniformly distributed for high values of $\Xi$. This is why ionic correlations enhance electroosmotic flow at small channel widths, as per Figs.~\ref{fig:plotbump} and \ref{fig:plotQ}. Such an effect appears in Fig.~\ref{fig4}, where fluid velocity profiles $\utilde(z)$ are shown for the same $\Xi$ and $d$ values as in Fig.~\ref{fig3}.

\begin{figure}
\includegraphics[width=\columnwidth]{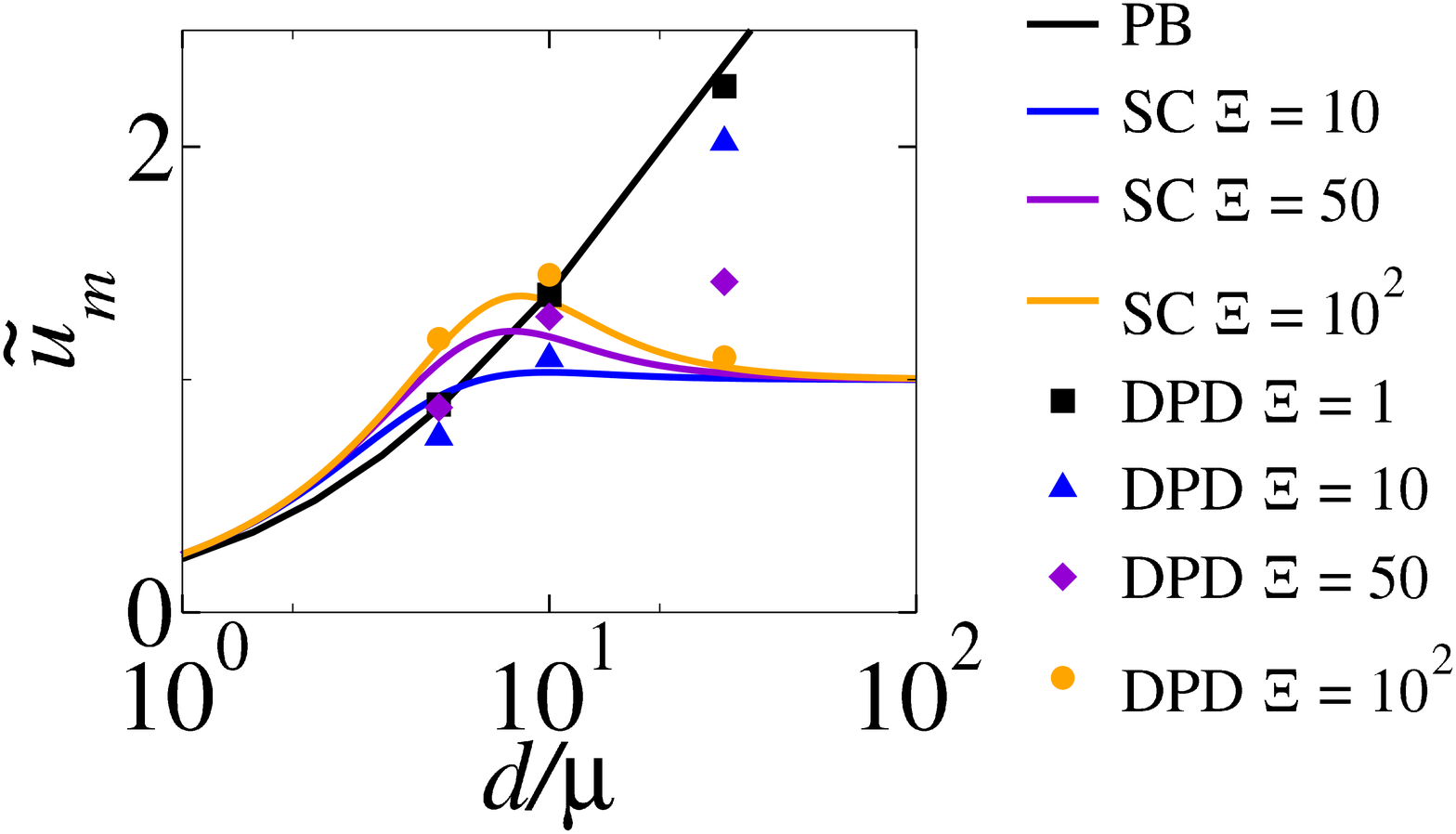}
\caption{Middle (maximum) fluid velocity as a function of separation $d/\mu$ for various coupling parameters $\Xi$, obtained with PB, Eq.~\eqref{eq:EOFPBasymptotic}, and SC theory ($\Xi>1$), Eq.~\eqref{eq:EOFsc}. 
Black squares, blue triangles, violet diamonds and orange circles represent DPD simulation data for MF, $\Xi=10$, $\Xi=50$ and $\Xi=100$, respectively.
}
\label{fig5}
\end{figure}

\begin{figure}
\includegraphics[width=\columnwidth]{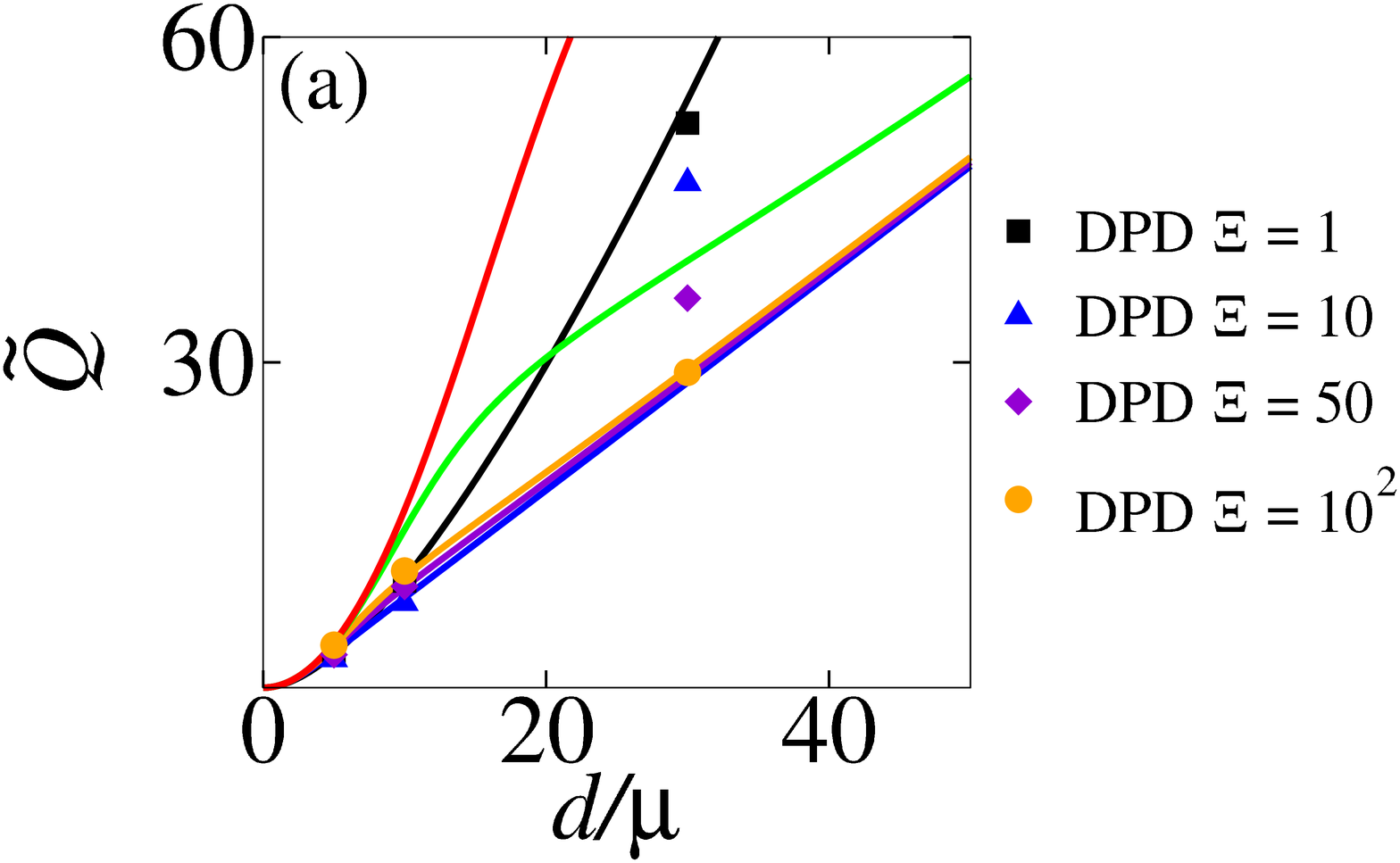}

\includegraphics[width=\columnwidth]{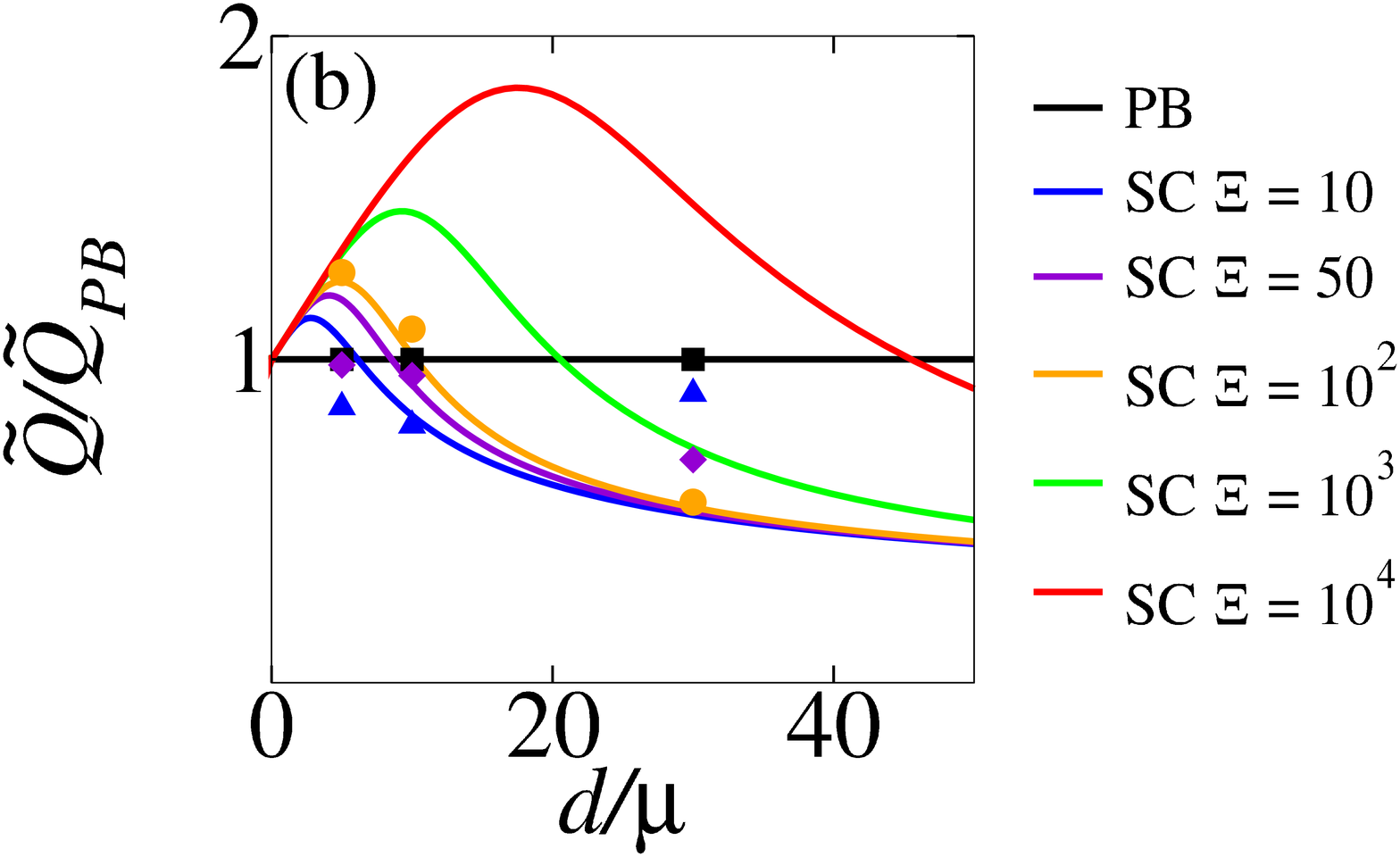}
\caption{(a) Integrated velocity profile as a function of separation $d/\mu$ for various coupling parameters $\Xi$. Symbols represent DPD simulations data. 
(b) Same quantity, normalised by its MF (PB) limiting law.
The color code is the same in both panels; symbols are for DPD simulations and lines for the theory.}
\label{fig6}
\end{figure}

We summarise these observations by analyzing the two scalar quantities we have focused on in Sec.~\ref{sec:Theory}: the peak velocity $\umax$ and the volume flow rate $\Qtilde$. These are represented, together with their analytical MF and SC predictions, in Figs.~\ref{fig5} and \ref{fig6}a. The theoretical curves are the same as in Figs.~\ref{fig:plotbump} and \ref{fig:plotQ}, respectively.

In general, the higher the coupling, the better our theory is at describing simulation results: in Fig.~\ref{fig3} and Fig.~\ref{fig4}, agreement with the SC curves increases from top to bottom, gradually subtracting credence to the MF results. By the time $\Xi$ reaches 100, the SC curves practically coincide with simulation results. However, it is worth noticing that the accuracy of the SC theory depends not only on $\Xi$, but also on $d$: at $\Xi=10$, for instance, the velocity profile deviates from the SC theory one only at $d=30\mu$ (Fig.~\ref{fig4}, second row).
Additional insight comes from the comparison of $\Xi=50$ to $\Xi=100$, at $d=30\mu$, in Figs.~\ref{fig3} and \ref{fig4}: for $\Xi=50$, the density only deviates from the SC curve in the middle of the channel -- when its value is hundreds of times smaller than close to the walls -- but this causes a conspicuous deviation in the velocity profile; for $\Xi=100$, at the same $d/\mu$, the agreement with SC is complete. This phenomenon is due to the high sensitivity of the electroosmotic flow to the decaying behaviour of the ionic density away from the charged surfaces (algebraic for MF, exponential for SC).
The question of a potential crossover to MF
and in particular of the recovery of
an algebraic tail,
for arbitrary coupling and very large separations, is presumably not relevant for the values of $d$ we probe here~\cite{Santangelo2006, Palaia2018}.



\section{Discussion and conclusion}
\label{sec:Conclusion}

Quantifying the role of ionic correlations in charged solutions is a challenging task. While static problems have been tackled through continuous~\cite{Netz2001, Santangelo2006, Bazant2011,Naji2013} and discrete~\cite{RoBl96,Samaj2011,Samaj2018a} approaches, comparatively little effort has been devoted to understanding the 
signature of beyond mean-field (MF) correlations on 
electrokinetics~\cite{Storey2012}. 
We have addressed here this question within planar electrosmosis, for a salt-free system.
The theory is simple and gives closed formulas for the relevant observables (e.g.~the velocity profile) for any $\Xi$, making use of no fitting parameter and no arbitrary correlation length scale (a common feature of many functional-based theories). Our findings are backed up by DPD. These simulations establish the regimes in which the theory is successful and those, limited to moderately small coupling parameters, in which MF is more accurate. 

We showed that ionic correlations can have two effects on electroosmotic properties, as compared to the MF expectation. On the one hand, they can boost the flow: this happens in the regime where they promote ionic delocalisation. Indeed, a uniform ionic distribution corresponds to a stronger flow. A correlated system exhibits an approximately uniform distribution up to inter-plane separations of order $\mu\Xi^{1/4}$, where $\mu$ is the Gouy-Chapman length; such delocalisation of the ions is more efficient than in MF theory and happens on a wider range of separations ($\mu\Xi^{1/4}\gg \mu$). On the other hand, correlations decrease the flow at much larger separations: in this regime, correlated ions exhibit a short-range distribution as a function of distance from the wall, and are therefore more confined than in an uncorrelated system, where the distribution is long-range.

In systems where the solvent has a high dielectric permittivity, the Gouy-Chapman length $\mu$ is usually very small under strong coupling regime, because of the large surface charge required for a large $\Xi$. The separations at which electroosmotic boost would be possible are rather small and interfere with molecular scales, so that, in these systems, it might only be possible to observe flow suppression. In low-permittivity solvents, often apolar, fewer systems reach surface charges for which $\Xi$ is large enough, but, in such cases, $\mu$ is larger. The left part of the $Q(d)$ diagram (Figs.~\ref{fig:plotQ} or \ref{fig6}b for instance) becomes then meaningful. In these systems, flow boost could be observable at distances of the order of tens of nanometres, while flow suppression will be seen at larger distances. In this regard, it is fair to point out that recent works \cite{Mukhina2019, Schlaich2019, CementInPrep} highlight a decrease of the permittivity of water under nanometric confinement: this might open up the boosted-flow regime to water based systems.

Having identified the signature of electrostatic correlations, it appears that electroosmotic flow may be used to probe the coupling parameter $\Xi$ in salt-free systems. Although well defined {\em a priori}, this quantity can indeed be elusive from an experimental point of view, through the measure of the wall surface charge. To this end, Figs. \ref{fig5} and \ref{fig6} may serve as a calibration reference to assess the importance of ionic correlations in an experimental system, or even in atomistic simulations.

Finally, perspectives include accounting for refined effects that may prove relevant under high surface charges, such as charge-induced thickening of the electrolyte, where the solution viscosity increases with charge density~\cite{Bazant2009}.

\section*{Conflicts of interest}
There are no conflicts to declare.

\section*{Acknowledgements}
We would like to thank Ladislav \v{S}amaj for useful discussions.
This work has received funding from the European Union’s Horizon 2020 research and innovation programme under the Marie Skłodowska-Curie grant agreement 674979-NANOTRANS, from the Brazilian Research Council (CNPq) under the grant 302720/2018-9 and from FAPERGS.






\bibliography{ref.bib} 

\end{document}